\documentclass[prd,nofootinbib,superscriptaddress,preprint]{revtex4}
\pdfoutput=1
\usepackage[T1]{fontenc}
\usepackage{amsmath,amssymb}
\usepackage{epsfig}
\usepackage{graphicx}
\usepackage[usenames,dvipsnames]{color}
\usepackage{slashed}
\usepackage[colorlinks,citecolor=blue]{hyperref}
\usepackage{pdfpages}
\usepackage{color}
\usepackage{subfig}

\newcommand{\Tr}[1]{\text{Tr}\big[#1\big]}

\usepackage{tikz-feynman}

\usepackage{tikzsymbols}
\usepackage{feynmp}
\begin{document}
\title{\Large A UV Complete Framework of Freeze-in Massive Particle Dark Matter}

\author{Anirban Biswas}
\email{anirban.biswas@iitg.ernet.in}
\affiliation{Department of Physics, Indian Institute of Technology
Guwahati, Assam 781039, India}
\author{Debasish Borah}
\email{dborah@iitg.ernet.in}
\affiliation{Department of Physics, Indian Institute of Technology
Guwahati, Assam 781039, India}
\author{Arnab Dasgupta}
\email{arnabdasgupta@protonmail.ch}
\affiliation{School of Liberal Arts, Seoul-Tech, Seoul 139-743, Korea}
\begin{abstract}
We propose a way to generate tiny couplings of freeze-in massive particle dark matter
with the Standard Model particles dynamically by considering an extension of the electroweak
gauge symmetry. The dark matter is considered to be a singlet under this extended gauge symmetry
which we have assumed to be the one in a very widely studied scenario called left-right symmetric
model. Several heavy particles, that can be thermally inaccessible in the early Universe due to
their masses being greater than the reheat temperature after inflation, can play the role of portals
between dark matter and Standard Model particles through one loop couplings. Due to
the loop suppression, one can generate the required non-thermal dark matter couplings without
any need of highly fine tuned Yukawa couplings beyond that of electron Yukawa with the
Standard Model like Higgs boson. We show that generic values of Yukawa couplings as large
as $\mathcal{O}(0.01)$ to $\mathcal{O}(1)$ can  keep the dark matter out of thermal equilibrium in the early Universe
and produce the correct relic abundance later through the freeze-in mechanism. Though the radiative couplings of dark matter are tiny as required by the freeze-in scenario, the associated
rich particle sector of the model can be probed at ongoing and near future experiments. The
allowed values of dark matter mass can remain in a wide range from keV to TeV order keeping
the possibilities of warm and cold dark matter equally possible.
\end{abstract}
\maketitle
\section{Introduction}
\label{intro}
In view of several astrophysical and cosmological evidences, the existence of non-baryonic form of matter, or the so called Dark Matter (DM) in large amount in the present Universe has become an irrefutable fact. Among these evidences, the galaxy cluster observations by Fritz Zwicky \cite{Zwicky:1933gu} back in 1933, observations of galaxy rotation curves in 1970's \cite{Rubin:1970zza}, the more recent observation of the bullet cluster \cite{Clowe:2006eq} and results from several satellite borne cosmology experiments like WMAP \cite{Hinshaw:2012aka} and Planck \cite{Ade:2015xua} are the most prominent ones. The precise measurements of the cosmology experiments reveal that more
than 80\% matter content of our Universe is in the form of this
non-baryonic or DM form. The amount of DM present in the Universe is often expressed by
a quantity $\Omega_{\rm DM} h^2$, which is called the relic density
of DM and it is the ratio of present mass density of DM by the critical
density of the Universe. The value of DM relic density at the present epoch
is $0.1172\leq\Omega_{\rm DM}h^2\leq0.1226$ at 67\% C.L. \cite{Ade:2015xua}.
Here $h = \dfrac{{\bf H}_0}{100\,{\rm km\,s^{-1}\,Mpc^{-1}}}$ is a parameter
of order unity while ${\bf H}_0$ being the present value of
the Hubble parameter.

In spite these astrophysical and cosmological evidences of DM, the information regarding
the constituents and origin of DM still remains unknown to us. One of the well
motivated and most studied scenario is to assume the thermal origin of DM
\cite{Srednicki:1988ce, Gondolo:1990dk},
where DM particles were produced thermally at the early Universe and
depending upon the mass and interaction strength, DM 
maintained both thermal and chemical equilibrium with
the plasma upto a certain temperature of the Universe.
Decoupling of DM from the thermal bath particles occurred
at around a temperature $T_f$, which is known as
the freeze-out temperature, where the interaction
rate of DM dropped below the expansion rate of the Universe
governed by the Hubble parameter $H$. Being
decoupled from the rest of the plasma these DM particles
becomes thermal relic whose density, after decoupling, is only affected by
the expansion of the Universe. The list of criteria a particle DM candidate
should fulfil rules out all the Standard Model (SM) particles from being DM candidates,
leading to several beyond Standard Model (BSM) proposals in the last few decades.
Most of the thermal DM candidates studied in the literature fall into a category
called weakly interacting massive particle (WIMP)
\cite{Jungman:1995df, Bertone:2004pz, Arcadi:2017kky},
which has mass in the range of few GeV to few TeV and weak scale couplings.
The interesting coincidence that a DM particle having mass and couplings around the
electroweak scale can give rise to the correct dark matter relic abundance is
often referred to as the \textit{WIMP Miracle}.

Now, if such type of particles whose interactions are of the order of electroweak
interactions really exist then we should expect their signatures in various DM direct detection experiments where
the recoil energies of detector nuclei scattered by DM particles
are being measured. However, after decades of running, direct detection
experiments are yet to observe any DM-nucleon scattering
\cite{Tan:2016zwf, Aprile:2017iyp, Akerib:2016vxi}. The absence of dark matter signals from the direct detection experiments have progressively lowered the exclusion curve in its mass-cross section plane. With such high precision measurements, the WIMP-nucleon cross section will soon overlap with the neutrino-nucleon cross section. Similar null results have been also reported by other direct search experiments like the large hadron collider (LHC) giving upper limits on DM interactions with the SM particles. A recent summary of collider searches for DM can be found in \cite{Kahlhoefer:2017dnp}. Although such null results could indicate a very constrained region of WIMP parameter space, they have also motivated the particle physics community to look for beyond the thermal WIMP paradigm where the interaction scale of DM particle can be much lower
than the scale of weak interaction i.e.\,\,DM may be more feebly
interacting than the thermal WIMP paradigm.     

One of the viable alternatives of WIMP paradigm, which may
be a possible reason of null results at various direct detection
experiments, is to consider the non-thermal origin of DM \cite{Hall:2009bx}. In this
scenario, the initial number density of DM in the early Universe is
negligible and it is assumed that the interaction strength of DM
with other particles in the thermal bath is so feeble that
it never reaches thermal equilibrium at any epoch in the early Universe. In this set up,
DM is mainly produced from the out of equilibrium decays
of some heavy particles in the plasma. It can also be produced
from the scatterings of bath particles, however if same couplings
are involved in both decay as well as scattering processes
then the former has the dominant contribution to DM relic density
over the latter one \cite{Hall:2009bx, Biswas:2016bfo, Biswas:2017tce}.
The production mechanism for non-thermal DM
is known as freeze-in and the candidates of non-thermal DM produced
via freeze-in are often classified into a group called
Freeze-in (Feebly interacting) massive particle (FIMP). For a recent review of this DM paradigm, please see \cite{Bernal:2017kxu}. Now, if the mother particle is in thermal
equilibrium with the bath then the maximum production of DM
occurs when the temperature of the Universe $T\simeq M_0$, the
mass of mother particle. Therefore, the non-thermality criterion
enforces the couplings to be extremely tiny via the following condition
$\left|\dfrac{\Gamma}{{\bf H}}\right|_{T\simeq M_0}<1$ \cite{Arcadi:2013aba}, where $\Gamma$
is the decay width. For the case of scattering, one has to replace $\Gamma$
by the interaction rate $n_{\rm eq}\, \langle \sigma {\rm v}\rangle$, $n_{eq}$
being the equilibrium number density of mother particle. 
These types of freeze-in scenarios are known as IR-freeze-in \cite{Yaguna:2011qn, Blennow:2013jba, Merle:2015oja, Shakya:2015xnx, Biswas:2016bfo,  Konig:2016dzg, Biswas:2016iyh, Biswas:2016yjr,  Bernal:2017kxu} where
DM production is dominated by the lowest possible temperature
at which it can occur i.e. $T\sim M_0$, since for $T<M_0$, the number
density of mother particle becomes Boltzmann suppressed. Here DM
interacts with the visible sector via renormalizable operators (dimension $d\leq 4$) only.
There may be a situation in IR-freeze-in, where mother particle itself
is out of thermal equilibrium and in such cases, first one has to calculate
the distribution function of mother particle considering its all possible production and
decay modes. This distribution function is necessary to compute
the non-thermal \footnote{Here we use the word non-thermal average
to distinguish it from the thermal average where the Maxwell-Boltzmann
distribution function is used \cite{Biswas:2016iyh}.}
averages of decay width and annihilation cross sections.
Once we know these quantities, the Boltzmann equation for the non-thermal
DM can be solved in terms of its comoving number density following
the usual procedure \cite{Konig:2016dzg, Biswas:2016iyh}.     

As mentioned earlier that to maintain a situation
where DM remains out of thermal equilibrium, one needs
extremely tiny couplings of DM with the particles in the plasma.
However, theoretically, the origin of such extremely low values
of couplings is in general, not obvious. One of the possible
explanation of such feeble interactions is to consider DM
to be connected to the visible sector via non-renormalizable
higher dimensional effective operators. This results in a
different type of freeze-in mechanism known as UV-freeze-in
\cite{Hall:2009bx, Elahi:2014fsa, McDonald:2015ljz},
where the comoving number density of DM is directly proportional
to reheat temperature $T_{RH}$ of the Universe and thus sensitive
to the early Universe cosmology. Another interesting way to generate
tiny dimensionless couplings is through the recently proposed clockwork mechanism
\cite{Kaplan:2015fuy, Giudice:2016yja} which has been recently explored in the
context of freeze-in DM by the authors of \cite{Kim:2017mtc, Kim:2018xsp}.
In this work, we try to explain the origin of such tiny couplings by considering a
renormalizable gauge extension of the SM where FIMP couplings with the rest of the particles can arise at radiative level, leading to the required suppression naturally. As an illustrative example, we consider a left-right symmetric gauge extension of SM where the FIMP candidate is a gauge singlet. However, at one loop level the gauge bosons can decay into the FIMP, with several particles going inside the loop. The particles in the loop do have sizeable couplings with FIMP, but that that does not lead to thermal production of FIMP DM if the corresponding scattering rates always remain smaller than the expansion rate of the Universe and their decay into FIMP are kinematically forbidden. However, if same couplings are involved in both scattering as well as the one loop decay, keeping scattering rates out of equilibrium typically makes the decay contribution very small. Another way is to consider these mediator particles to be too heavy to be produced at the end of inflation (having mass more than the reheat temperature). We adopt this approach without going into the details of specific inflationary models and find the predictions for the dark sector. Such an exercise can be carried out for simpler gauge extension of SM as well, but we perform it here for left-right symmetric model (LRSM) which has several other motivations.

Rest of the article is organised as follows. In section \ref{sec1}, we discuss our model followed by the details of the calculation of one loop vertex factors of dark matter interactions with the neutral gauge bosons in section \ref{loop}. In section \ref{sec:fimp} we discuss the details of dark matter calculations, results and then finally conclude in section \ref{sec:conc}
     
\section{The Model}
\label{sec1}

The LRSM is one of most highly motivated BSM frameworks which in its generic form \cite{Pati:1974yy, Mohapatra:1974gc, Senjanovic:1975rk, Mohapatra:1980qe, Deshpande:1990ip}, not only explains the origin of parity violation in weak interactions but also explains the origin of tiny neutrino masses naturally. The gauge symmetry group and the field content of the generic LRSM can also be embedded within grand unified theory (GUT) symmetry groups like $SO(10)$ providing a non-supersymmetric route to gauge coupling unification. The right handed fermions of the SM forms doublet under a new $SU(2)_R$ group in LRSM such that the theory remains parity symmetric at high energy. This necessitates the inclusion of the right handed neutrino as a part of the right handed lepton doublet. To be more appropriate, the  gauge symmetry of the Standard Model namely, $SU(3)_c \times SU(2)_L \times U(1)_Y$ is upgraded to $SU(3)_c \times SU(2)_L \times SU(2)_R \times U(1)_{B-L}$ such that the right handed fermions transform as doublets under $SU(2)_R$, making the theory left-right symmetric. The model also has an in-built discrete $\mathbb{Z}_2$ symmetry or D-parity which ensures the equality of couplings in $SU(2)_{L,R}$ sectors. The effective parity violating electroweak physics at low energy arises as a result of spontaneous breaking of the $SU(2)_R \times U(1)_{B-L} \times D$ to $U(1)_Y$ of the SM. 

The minimal LRSM however, does not contain a naturally stable DM candidate. One can of course realise a long lived keV right handed neutrino DM in these models. Such a scenario leading to warm dark matter scenarios has been investigated within LRSM in \cite{Nemevsek:2012cd, Bezrukov:2009th, Borah:2017hgt}. Due the presence of $SU(2)_R$ gauge interactions, such a right handed neutrino dark matter can be thermally produced in the early Universe, unlike in typical keV right handed neutrino DM models where non-thermal origin is required \cite{Adhikari:2016bei}. On the other hand, to have WIMP DM type realisation, the minimal LRSM can be extended by additional scalar or fermion multiplets in the spirit of minimal DM scenario \cite{Cirelli:2005uq,Garcia-Cely:2015dda,Cirelli:2015bda}. Such minimal dark matter scenario in LRSM has been studied recently by the authors of \cite{Heeck:2015qra,Garcia-Cely:2015quu}. In these models, the dark matter candidate is stabilised either by a $\mathbb{Z}_2 = (-1)^{B-L}$ subgroup of the $U(1)_{B-L}$ gauge symmetry or due to an accidental symmetry at the renormalisable level due to the absence of any renormalisable operator leading to dark matter decay \cite{Borah:2016hqn}. Some more studies on left-right dark matter also appeared in the recent works \cite{Borah:2016ees, Berlin:2016eem, Borah:2016lrl, Borah:2017leo, Borah:2017xgm}. The possibility of right handed neutrino dark matter in a different version of LRSM where the right handed lepton doublets do not contain the usual charged leptons, was also studied in the recent works \cite{Dev:2016qbd, Dev:2016xcp, Dev:2016qeb}.

In this work, we intend to have a purely non-thermal DM within LRSM. We therefore consider gauge singlet fermion $N$ as our DM candidate. We introduce an additional $\mathbb{Z}_2$ symmetry under which this new singlet fermion is odd and hence can be stable if it happens to be the lightest $Z_2$ odd particle. One can also consider scalar singlet DM, but scalars usually have quartic couplings with other scalars and it is often difficult to forbid them from symmetry arguments. We also introduce two copies of vector like fermion doublets $\psi$ and a pair of scalar doublets $H_{L,R}$ to the minimal LRSM. These additional fields play the role of generating interactions of the SM sector with the DM particle $N$ at radiative level, as we will see below.

\begin{table}
\begin{center}
\begin{tabular}{|c|c|}
\hline
Particles & $SU(3)_c \times SU(2)_L \times SU(2)_R \times U(1)_{B-L} \times \mathbb{Z}_2$   \\
\hline
$Q_L$ & $(3,2,1,\frac{1}{3}, +)$ \\
$Q_R$ & $(3,1,2,\frac{1}{3}, +)$  \\
$\ell_L$ & $(1,2,1,-1, +)$  \\
$\ell_R$ & $(1,1,2,-1, +)$ \\
$\psi$ & $(1,2,1,-1, -)$ \\
$\psi^{\prime}$ & $(1,1,2,-1, -)$ \\
$N$ & $(1,1,1,0, -)$ \\
\hline
\end{tabular}
\end{center}
\caption{Fermion content of the model}
\label{tab1}
\end{table}
\begin{table}
\begin{center}
\begin{tabular}{|c|c|}
\hline
Particles & $SU(3)_c \times SU(2)_L \times SU(2)_R \times U(1)_{B-L}\times \mathbb{Z}_2$   \\
\hline
$\Phi$ & $(1,2,2,0,+)$ \\
$\Delta_L$ & $(1,3,1,2,+)$ \\
$\Delta_R$ & $(1,1,3,2,+)$ \\
$H_L$ & $(1,2,1,-1,+)$ \\
$H_R$ & $(1,1,2,-1,+)$ \\
$ \sigma  $ & $(1,1,1,0,+)$\\
\hline
\end{tabular}
\end{center}
\caption{Scalar content of the model}
\label{tab2}
\end{table}
The relevant Yukawa couplings for the Standard Model fermion masses can be written as 
\begin{align}
{\cal L}^{SM}_{Y} &= y_{ij} \bar{\ell}_{iL} \Phi \ell_{jR}+ y^\prime_{ij} \bar{\ell}_{iL}
\tilde{\Phi} \ell_{jR} +Y_{ij} \bar{Q}_{iL} \Phi Q_{jR}+ Y^\prime_{ij} \bar{Q}_{iL}
\tilde{\Phi} Q_{jR}
\nonumber \\
& \qquad + \frac{1}{2}(f_L)_{ij} \ell_{iL}^T \ C \ i \sigma_2 
\Delta_L \ell_{jL}+\frac{1}{2}(f_R)_{ij} \ell_{iR}^T \ C \ i \sigma_2 \Delta_R \ell_{jR}+\text{H.c.}
\label{treeY}
\end{align}
where $\tilde{\Phi} = \tau_2 \Phi^* \tau_2$, $C$ is the charge conjugation operator and the indices $i, j = 1, 2, 3$ correspond to the three generations of fermions. The Yukawa couplings involving the new fermions can be written as
\begin{align}
{\cal L}^{new}_{Y} &=Y_{\psi} \bar{\psi} \tilde{H_L} N +  Y_{\psi^\prime} \bar{\psi^{\prime}} \tilde{H_R} N + M \bar{\psi}\psi + M^{\prime} \bar{\psi^{\prime}}\psi^{\prime} +f_{\psi} \sigma(\bar{\psi}\psi- \bar{\psi^{\prime}}\psi^{\prime}) \nonumber \\
& + Y_{\phi} \bar{\psi} \Phi \psi^{\prime}+ Y^{\prime}_{\phi} \bar{\psi} \tilde{\Phi} \psi^{\prime}
\label{yukawa1} 
\end{align}

The details of the scalar potential is given in appendix \ref{appen1}. At a very high energy scale, the parity odd singlet $\sigma$ can acquire a vev to break D-parity spontaneously while the neutral component of $\Delta_R$ acquires a non-zero vev at a later stage to break the gauge symmetry of the LRSM into that of the SM which then finally gets broken down to the $U(1)_{\rm em}$ of electromagnetism by the vev of the neutral component of Higgs bidoublet $\Phi$. Thus, the symmetry breaking chain is 
$$ SU(2)_L \times SU(2)_R \times U(1)_{B-L} \times D \quad \underrightarrow{\langle \sigma \rangle} \quad SU(2)_L \times SU(2)_R \times U(1)_{B-L} \quad \underrightarrow{\langle
\Delta_R \rangle} \quad SU(2)_L\times U(1)_Y$$
$$ SU(2)_L\times U(1)_Y \quad \underrightarrow{\langle \Phi \rangle} \quad U(1)_{\rm em}$$
Denoting the vev of the neutral components of the bidoublet as $k_{1,2}$ and that of triplet $\Delta_R$ as $v_R$, the gauge boson masses after spontaneous symmetry breaking can be written as
$$ M^2_{W_L} = \frac{g^2}{4} k^2_1, \;\;\; M^2_{W_R} = \frac{g^2}{2}v^2_R ~, $$
$$ M^2_{Z_L} =  \frac{g^2 k^2_1}{4\cos^2{\theta_w}} \left ( 1-\frac{\cos^2{2\theta_w}}{2\cos^4{\theta_w}}\frac{k^2_1}{v^2_R} \right), \;\;\; M^2_{Z_R} = \frac{g^2 v^2_R \cos^2{\theta_w}}{\cos{2\theta_w}}~, $$
where $\theta_w$ is the Weinberg angle. The neutral components of the other scalar fields $H_{L,R}$ do not acquire any vev. However, the neutral component of the scalar triplet $\Delta_L$ can acquire a tiny but non-zero induced vev after the electroweak symmetry breaking as
\begin{equation}
v_{L}=\gamma \frac{M^{2}_{W_L}}{v_{R}}~,
\label{deltaLvev}
\end{equation}
with $M_{W_L}\sim 80.4$ GeV being the weak boson mass and $\gamma$ is a function of various couplings in the scalar potential. The bidoublet also gives rise to non-zero $W_L-W_R$ mixing parameterised by $\xi$ as
\begin{equation}
\tan{2\xi} = \frac{2k_1 k_2}{v^2_R-v^2_L}\;,
\end{equation}
which is constrained to be $\xi \leq 7.7 \times 10^{-4}$ \cite{ATLAS:2012ak, CMS:2012zv}. 

It should be noted that, our scenario can work even without the D-parity odd singlet scalar. In such a case, the parameters of the left and right sectors are equal until the $SU(2)_R \times U(1)_{B-L}$ symmetry breaking scale.

\section{Decay of $Z_{L,R}$ into FIMP}
\label{loop}

\begin{figure}[ht]
    \centering
\includegraphics[scale=0.75]{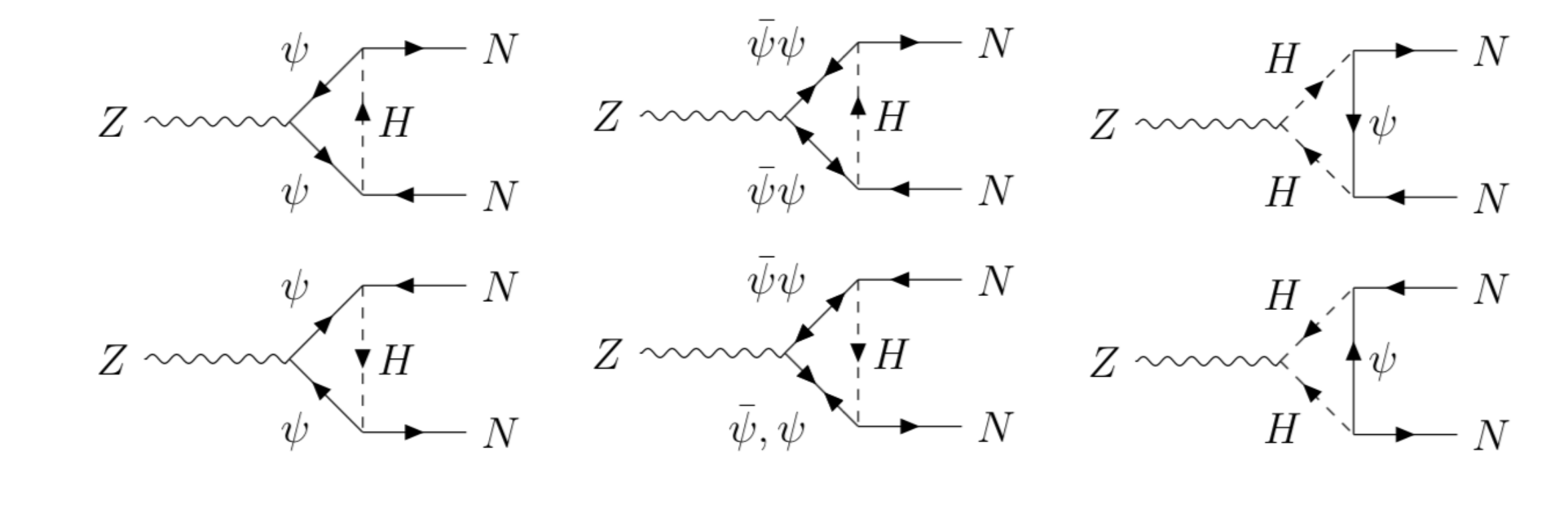}
    \caption{The relevant decay diagrams for the $Z \rightarrow N N$ in two-spinor notation. Here for $Z=Z_L(Z_R)$,
    $f=\psi(\psi^\prime)$ and $H=H_L(H_R)$.}
    \label{fig:Z_decay}
\end{figure}

The decay of $Z_{L,R}$ to the dark matter can occur at one loop level,
the Feynman diagrams for which are shown in Fig.\,\,\ref{fig:Z_decay}
: \\
The Lagrangian for the decay process is $-i\mathcal{A}\bar{N}\gamma^\mu\gamma_5 Z_{\mu L, R} N$
where the loop factor $\mathcal{A}$ is given as:
\begin{align}
   \mathcal{A} &= \frac{gY^*Y}{16\pi^2}\left[1 + \frac{m^2_{H_{L,R}}}{M^2_{Z_{L,R}}}\ln\left[\frac{2m^2_{H_{L,R}}-M^2_{Z_{L,R}} + \sqrt{M^4_{Z_{L,R}} - 4 m^2_{H_{L,R}}M^2_{Z_{L,R}}}}{2m^2_{H_{L,R}}}\right]^2\right]\,\,,
\label{loopfimp}    
\end{align}
if we take the limit $m^2_{H_{L,R}}\ll M^2_{Z_{L,R}}$ and use the parametrisation $y=\frac{m^2_{H_{L,R}}}{M^2_{Z_{L,R}}}$, the loop factor can be written as
\begin{align*}
    \mathcal{A} &= \frac{gY^*Y}{16\pi^2}\left[1 + y\ln[y]^2 - \pi^2y + 2i\pi \ln y + \mathcal{O}(y^2)\right].
\end{align*}
On the other hand, going to the other limit $m^2_{H_{L,R}}\gg M^2_{Z_{L,R}}$ and parametrising $x=\frac{M^2_{Z_{L,R}}}{m^2_{H_{L,R}}}$ we can write
\begin{align*}
    \mathcal{A} &= \frac{gY^*Y}{16\pi^2}\left[-\frac{x}{12} + \mathcal{O}(x^{3/2})\right]\,\,.
\end{align*}
For further details about the loop diagram computation, please refer to Appendix \ref{loopcalc}.
\begin{figure}
\includegraphics[width=0.6\textwidth]{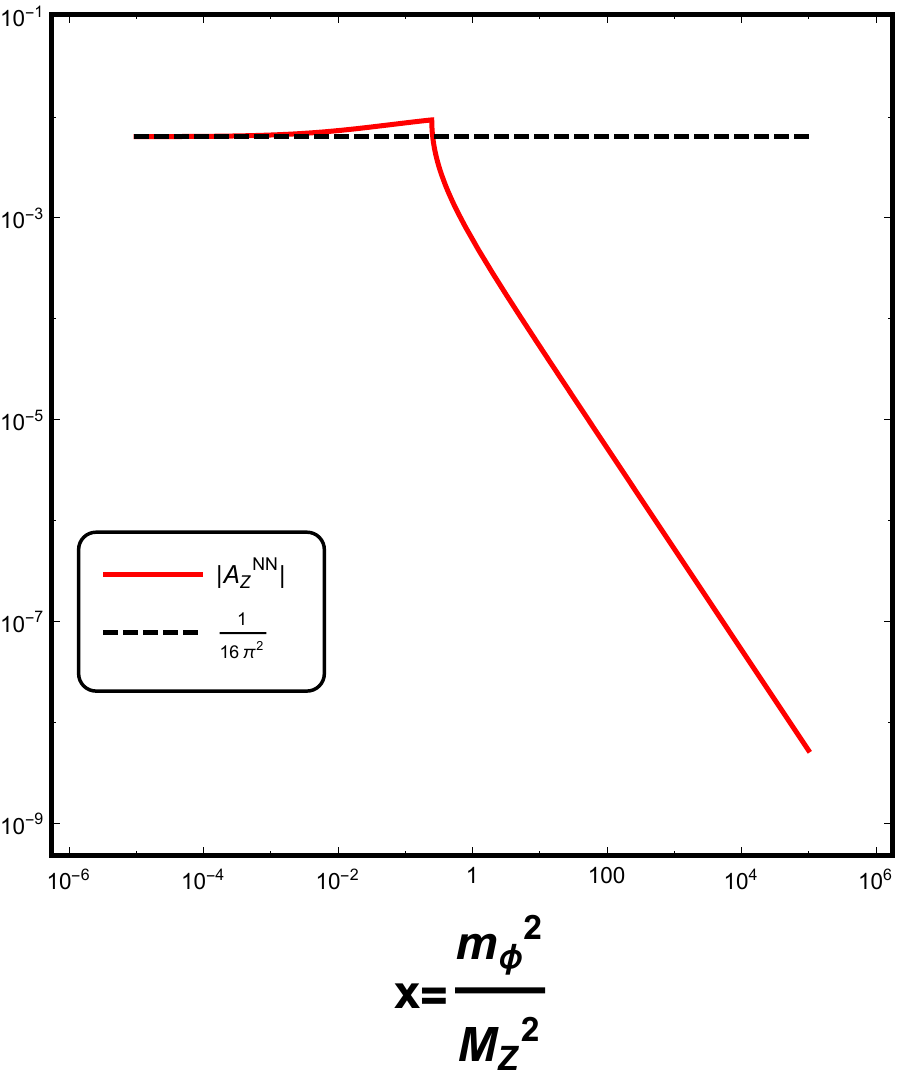}
\caption{Form factor for the one loop decay shown in figure \ref{fig:Z_decay}.}
\label{formf}
\end{figure}
In the above expression, all the masses of the particles inside the loop are taken to be same that is, $M_{\psi, \psi'}=M_{H_{L,R}}=m_{\phi}$. Also, the Yukawa couplings are denoted by a generalised notation $Y$ which for $Z_L$ as mother particle corresponds to $Y_{\psi}$ and for $Z_R$ as mother particle corresponds to $Y_{\psi^\prime}$ shown in the Lagrangian in Eq. \eqref{yukawa1}.

Since the particles inside loop have bare mass terms and hence can decouple, the form factor becomes very small for large mass for loop particles and in fact vanishes for $m_{\phi} \rightarrow \infty$. This behaviour can be seen from the plot of form factor as a function of $x = m^2_{\phi}/M^2_Z$ shown in Fig. \ref{formf}. To show the dependence on loop particle masses alone, here we assume the Yukawa couplings to be unity. Suitable tuning of Yukawa couplings can help us to achieve the required vertex factor for freeze-in dark matter even if the loop particles are few times heavier than the decaying one.

\section{Relic Density Calculation}
\label{sec:fimp}
\subsection{$N$ as a natural FIMP candidate}
\begin{figure}
\centering
\includegraphics[height=7cm,width=8.1cm,angle=-0]{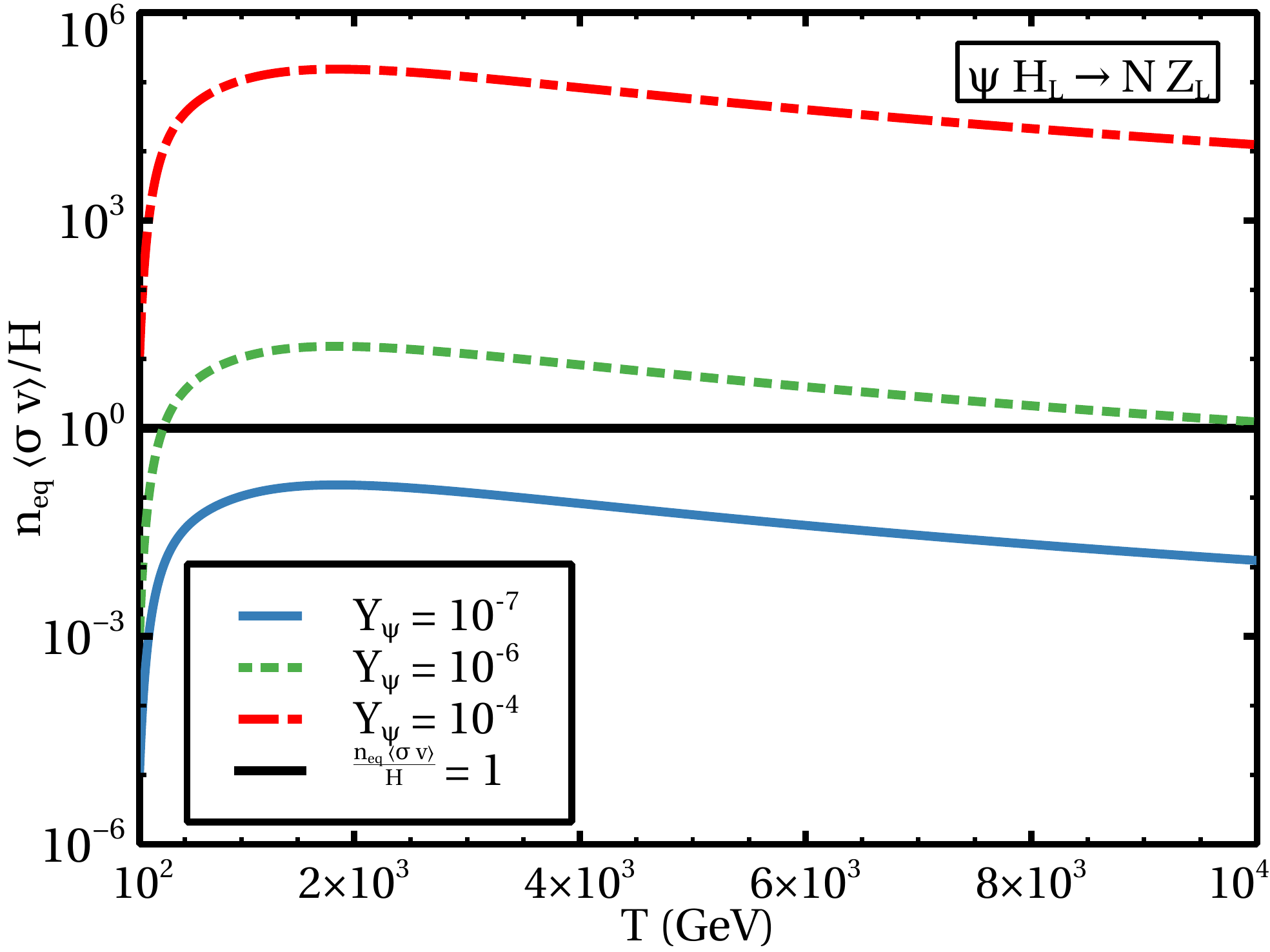}
\includegraphics[height=7cm,width=8.1cm,angle=-0]{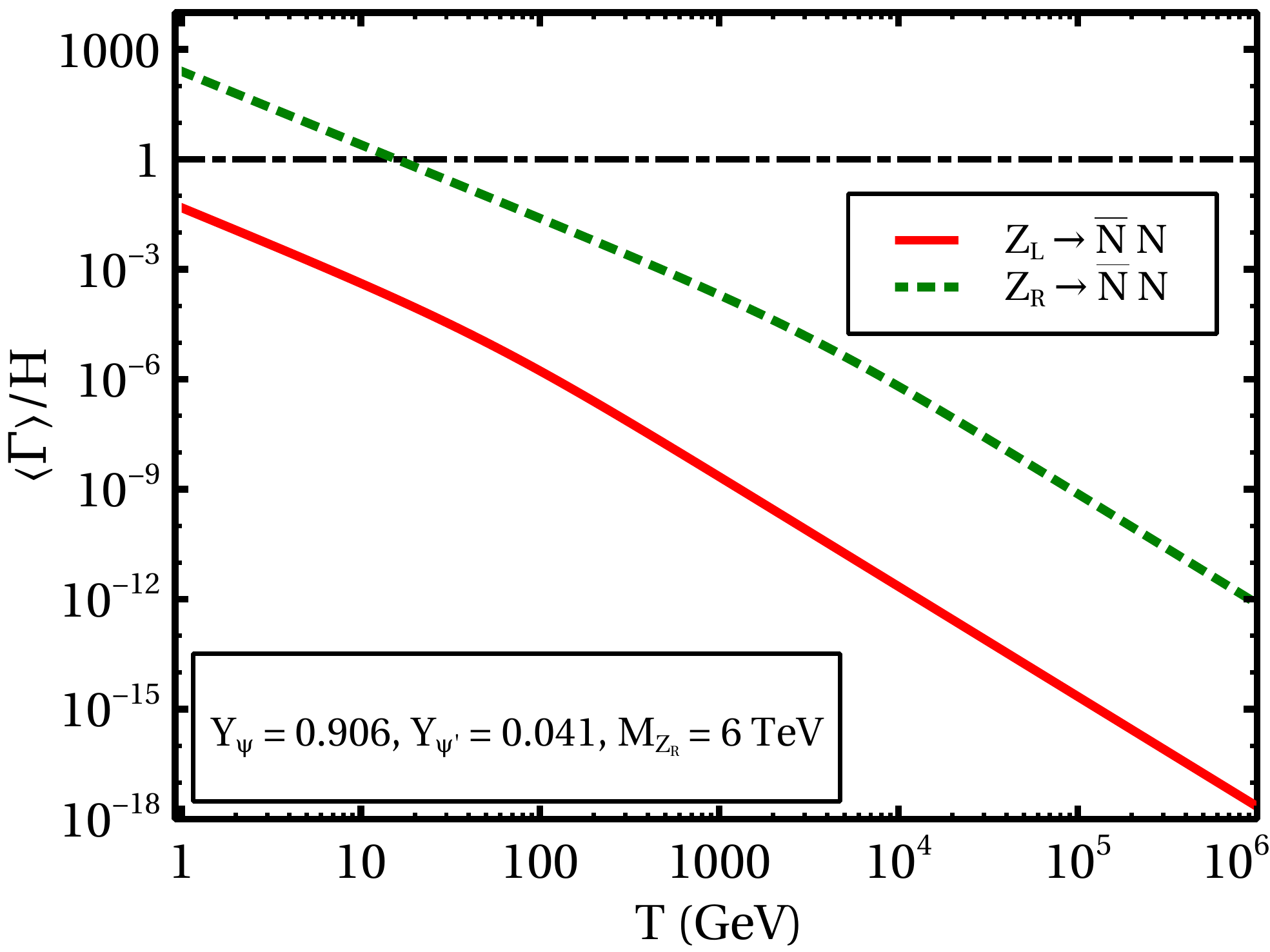}
\caption{Variation of ratio between interaction rate and Hubble expansion
rate with temperature of the Universe. Left panel: plot for the
scattering $\psi H_L \rightarrow N Z_L$. Right panel:
plot for the decay modes $Z_{L,\,R}\rightarrow \bar{N}N$.}
\label{Fig:rate}
\end{figure}

In this model, as mentioned earlier, the fermion $N$
is completely singlet under the left-right symmetry group
$SU(2)_{L} \times SU(2)_{R} \times U(1)_{B-L}$ and it
has an odd $\mathbb{Z}_2$ parity. However, being the
lightest $\mathbb{Z}_2$-odd particle, all the $\mathbb{Z}_2$
parity conserving decay modes of $N$ are kinematically
forbidden, making $N$ absolutely stable over the cosmological
time scale. Therefore, we consider $N$ as a suitable dark matter
candidate in this work. Apart from being the lightest $\mathbb{Z}_2$-odd particle,
$N$ also fits into our desired FIMP scenario. This is due to the fact that,
in the present model, all the portal interactions of $N$ via $Z_{L,\,R}$
and $\Phi$ with the Standard Model particles are one loop suppressed.
This naturally makes $N$ very feebly interacting with the thermal bath and as
a result $N$ remains out of thermal equilibrium. Although $N$
can thermalise with the plasma via t-channel scattering processes
like $H_{L\,(R)}\,H_{L\,(R)}\rightarrow \overline{N}\, N$,
$\overline{\psi^{(\prime)}}\, \psi^{(\prime)}\rightarrow \overline{N} N$,
$\psi^{(\prime)} H_{L\,(R)}\rightarrow\,N\,Z_{L\,(R)}$,
$\psi^{(\prime)} H^{+}_{L\,(R)}\rightarrow\,N\,W^{+}_{L\,(R)}$ etc,
where the two $N$ final states are suppressed by $Y^4_{\psi\,(\psi^{\prime})}$
while the scattering processes with single $N$ in the final state
are proportional to $Y^2_{\psi\,(\psi^{\prime})}$. Since, the Yukawa couplings
are $Y_{\psi}\,Y_{\psi^{\prime}}< 1$ in general, the dominant contribution arises
from the scattering processes with single $N$ in the final state.
In the left panel of Fig.\,\,\ref{Fig:rate} we demonstrate the variation
of ratio between the interaction rate $n_{\rm eq}\langle \sigma v\rangle$
for the scattering process $\psi H_L \rightarrow N Z_L$
and the Hubble expansion rate ${\bf H}$ 
with the temperature of the Universe T. In this plot,
we have considered the masses of all the components
of $H_L$ and $\psi$ to be equal and we have kept them
fixed at 1 TeV. From this plot one can easily notice
that the ratio $\dfrac{n_{\rm eq}\langle \sigma v\rangle}{\bf H}$
is maximum when $T\sim 1$ TeV, same as the mass of incoming particles.
After that, $\dfrac{n_{\rm eq}\langle \sigma v\rangle}{\bf H}$
reduces with T as the equilibrium number density becomes
exponentially suppressed for $T<1$ TeV. 
We have plotted the variation of $\dfrac{n_{\rm eq}\langle \sigma v\rangle}{\bf H}$
with respect to $T$ for three different values of Yukawa coupling
e.g. $Y_{\psi}=10^{-4}$, $10^{-6}$ and $10^{-7}$ respectively.
We have found that for $Y_{\psi}\geq 10^{-6}$, our DM
candidate remains in thermal equilibrium with the plasma
through the scattering $\psi^{(\prime)} H_{L\,(R)}\rightarrow\,N\,Z_{L\,(R)}$,
as the interaction rate exceeds the Hubble expansion rate
for the considered range of Temperature
$10^{2}\,\,{\rm GeV} \leq T \leq 10^{4}\,\,{\rm GeV}$.
However, as we reduce the Yukawa coupling further, the corresponding
interaction rate also decreases ($\propto Y^2_{\psi}$) and
we have found that for $Y_{\psi}\leq 10^{-7}$, $N$ never
thermalises with the bath particles. Therefore, the
non-thermality condition (${n_{\rm eq}\langle \sigma v\rangle}/{\bf H}<1$)
demands $Y_{\psi}, Y_{\psi^{\prime}}\leq 10^{-7}$
\footnote{The similar bound on $Y_{\psi^{\prime}}$ is coming from
scattering involving $\psi^{\prime}$, $H_R$, $N$ and $Z_R$.}.

Nevertheless, one can still consider $N$
as a non-thermal dark matter candidate without assuming
such a low value for the Yukawa couplings. This is in fact, the prime motivation
of this article, to realise FIMP dark matter without highly fine-tuned dimensionless couplings.
In that case, we have to consider a scenario with low reheat temperature
\cite{deSalas:2015glj} of the Universe so that these particles
$H_{L,R}, \psi, \psi^{\prime}$ were not thermally produced in
the early Universe, which actually prevents $N$ to thermalise via scattering with $H_{L}$, $H_R$,
$\psi$ and $\psi^{\prime}$. For example, the authors of \cite{Mambrini:2013iaa} considered such heavy mediators having mass greater than the reheat temperature, but in a different dark matter scenario. Since the masses of these particles appear as bare mass terms in the Lagrangian and hence they do not depend upon the scale of symmetry breaking, we can push them high to any scale above the scale of reheating. Therefore, we do not really have to rely upon very specific inflationary models where low reheat temperature occurs. Such a setup leads to a situation where $N$ can `talk' to other particles in the
thermal bath only through the one loop suppressed
portals like $Z_L$, $Z_R$ and $\Phi$. In the right
panel of Fig.\,\,\ref{Fig:rate}, we plot the
ratio of thermally averaged partial decay widths of $Z_{L,\,R}$
into two $N$'s and ${\bf H}$ with $T$. Here, green
dashed line represents the variation $\dfrac{\langle\Gamma\rangle}{\bf H}$
with $T$ for the right handed neutral gauge boson $Z_R$.
In this plot, we have considered $M_{Z_R} = 6$ TeV and
the corresponding Yukawa coupling $Y_{\psi^\prime} = 0.041$
while the similar plot for the left handed neutral gauge
boson $Z_L$ has been drawn for $Y_{\psi}= 0.906$
and this is represented by the red solid line. From this figure,
it is evident that for such moderately large Yukawa couplings
(significantly larger compared to those required for scattering
to be out of equilibrium) the decay rates of
$Z_L\rightarrow \bar{N}N$, $Z_R\rightarrow \bar{N}N$
always lie below the Hubble expansion rate ${\bf H}$, thus
maintaining the non-thermality criteria of $N$.
Although, for $Z_R$, the corresponding decay rate exceeds
${\bf H}$ when $T<10$ GeV, however at such low temperature
the number density of $Z_R$ with mass 6 TeV becomes
exponentially suppressed and hence does have a
very little impact on $N$. In fact, $\dfrac{\langle\Gamma\rangle}{\bf H}$
for both $Z_L$ and $Z_{R}$ lie well below the expansion
rate at $T\sim M_{Z_i}$ ($i=L$, $R$) where the maximum
production of $N$ from the decay of $Z_i$ occurs.    
Therefore, $N$ remains out of thermal equilibrium
for moderately large values of Yukawa
couplings ($\lesssim\mathcal{O}(1)$) and
can be a candidate for FIMP accordingly. Note that the Yukawa couplings
considered above are not fine tuned ones compared to those required in
generic FIMP models (IR Freeze-in scenarios).
\subsection{The Boltzmann equation}
The Boltzmann equation which governs
the evaluation of comoving number density
(ratio of number density to entropy density) of a FIMP is given by  

\begin{eqnarray}
\dfrac{dY_{N}}{dz} &=& \dfrac{2 M_{\rm Pl}}{1.66 M_{sc}^{2}}
\dfrac{z \sqrt{g_{\star}(z)}}{g_{s}(z)}\,\,\Bigg[\sum_{\chi\,=\,Z_{L},
\,Z_{R},\,\Phi}\langle\Gamma_{\chi\rightarrow  \bar{N} N}
\rangle(Y_{\chi}^{\rm eq} - Y_{N})\Bigg] \nonumber \\
&&+\dfrac{4 \pi^{2}}{45} \dfrac{M_{pl} M_{\rm sc}}{1.66}
\dfrac{\sqrt{g_{\star}(z)}}
{z^{2}}\,\,\Bigg[\sum_{p\,=\,\text{SM\,fermions}}
\langle\sigma {\rm v}_{p\bar{p}\rightarrow 
\bar{N}N}\rangle \{(Y_{p}^{\rm eq})^2 - Y_{N}^{2} \} \Bigg]\,.
\label{befimp}
\end{eqnarray}

where $z=\dfrac{M_{\rm sc}}{T}$, is a dimensionless variable while
$M_{\rm sc}$ is some arbitrary mass scale which we choose equal to
the mass of $Z_L$ and $M_{\rm Pl}$ is the Planck mass. Moreover, $g_s(z)$
is the number of effective degrees of freedom associated to the
entropy density of the Universe and the quantity $g_{\star}(z)$
is defined as
\begin{eqnarray}
\sqrt{g_{\star}(z)} = \dfrac{g_{\rm s}(z)}
{\sqrt{g_{\rho}(z)}}\,\left(1 -\dfrac{1}{3}
\dfrac{{\rm d}\,{\rm ln}\,g_{\rm s}(z)}{{\rm d} \,{\rm ln} z}\right)\,. 
\label{gstarz}
\end{eqnarray}
Here, $g_{\rho}(z)$ denotes the effective number of degrees
of freedom related to the energy density of the Universe at
$z=\dfrac{M_{\rm sc}}{T}$. The first term in the right hand side of the
above Boltzmann equation \eqref{befimp} represents the production of our dark matter
candidate $N$ from the decays of $Z_L$, $Z_R$ and bi-doublet
$\Phi$. The quantity $Y^{\rm eq}_{\chi}$ is the equilibrium comoving
number density of the species $\chi$ ($\chi=Z_L$, $Z_R$, $\Phi$)
and in this work, except the dark matter candidate $N$, we consider
Maxwell-Boltzmann distribution function for all the other particles
which are in thermal equilibrium. As we have mentioned earlier, the
production processes of $N$ from the decays of $Z_L$, $Z_R$ and
$\Phi$ are one loop suppressed. The Feynman diagrams of these
processes are shown in Fig.\,\ref{fig:Z_decay} and the corresponding one loop vertices are given in Eq.\,\eqref{loopfimp}.
Using the expressions of one loop vertices one can easily
compute the thermally averaged decay width for the processes
$Z_{L,\,R}\rightarrow\bar{N}N$ and expressions are given by,
\begin{eqnarray}
\langle \Gamma_{{Z_{L,\,R}}\,\rightarrow\bar{N}N} \rangle
&=& \Gamma_{{Z_{L,\,R}}\,\rightarrow\bar{N}N}
\dfrac{{\rm K}_1\left(\frac{M_{Z_{L,\,R}}}{T}\right)}
{{\rm K}_2\left(\frac{M_{Z_{L,\,R}}}{T}\right)}\,,\\
\Gamma_{{Z_{L,\,R}}\,\rightarrow\bar{N}N} &=&
\dfrac{\left|\mathcal{A}^{NN}_{Z_{L,\,R}}\right|^2\,M_{Z_{L,\,R}}}{24\,\pi}
\left(1-\dfrac{4\,M^2_N}{M^2_{Z_{L,\,R}}}\right)^{3/2}\,.
\label{gammaZLR}
\end{eqnarray}
Where, ${\rm K}_n\left(\frac{M_{Z_{L,\,R}}}{T}\right)$ is the
$n$-th order modified Bessel function of second kind.
The second term in the right hand side of the Boltzmann equation
represents the contributions coming from the annihilations
of SM particles to the production processes of $N$.
Since $N$ is a singlet under the left-right symmetry group,
the interactions of $N$ with the SM particles are possible
only through the portal interactions by $Z_L$, $Z_R$ and $\Phi$ beyond tree level.
One could have a tree level coupling like $\bar{\ell}_{L} \tilde{H}_L N$
if the additional discrete symmetry $\mathbb{Z}_2$ was not in place.
The other tree level couplings involving $H_{L,R}, \psi, \psi^{\prime}$
and $N$ will not play a role in the production of $N$ if these heavy
incoming particles were not thermally produced in the early Universe
due to their masses being heavier than the reheat temperature after
inflation, as argued previously.

Hence, the contributions of such processes are sub-dominant
compared to that from the decays of $Z_L$, $Z_R$, $\Phi$ as the couplings
between our FIMP dark matter $N$ and these mediator particles
are one loop suppressed. In our calculations, we consider only the decay
of the neutral gauge bosons $Z_L$, $Z_R$ as they are more likely to be
dominant due to the presence of gauge couplings in one of the vertices
of the one loop diagram. For $\Phi$ decay diagram, one has more freedom
in choosing another Yukawa coupling and hence that contribution can
remain suppressed compared to the gauge boson ones.

\begin{figure}
\includegraphics[scale=0.4]{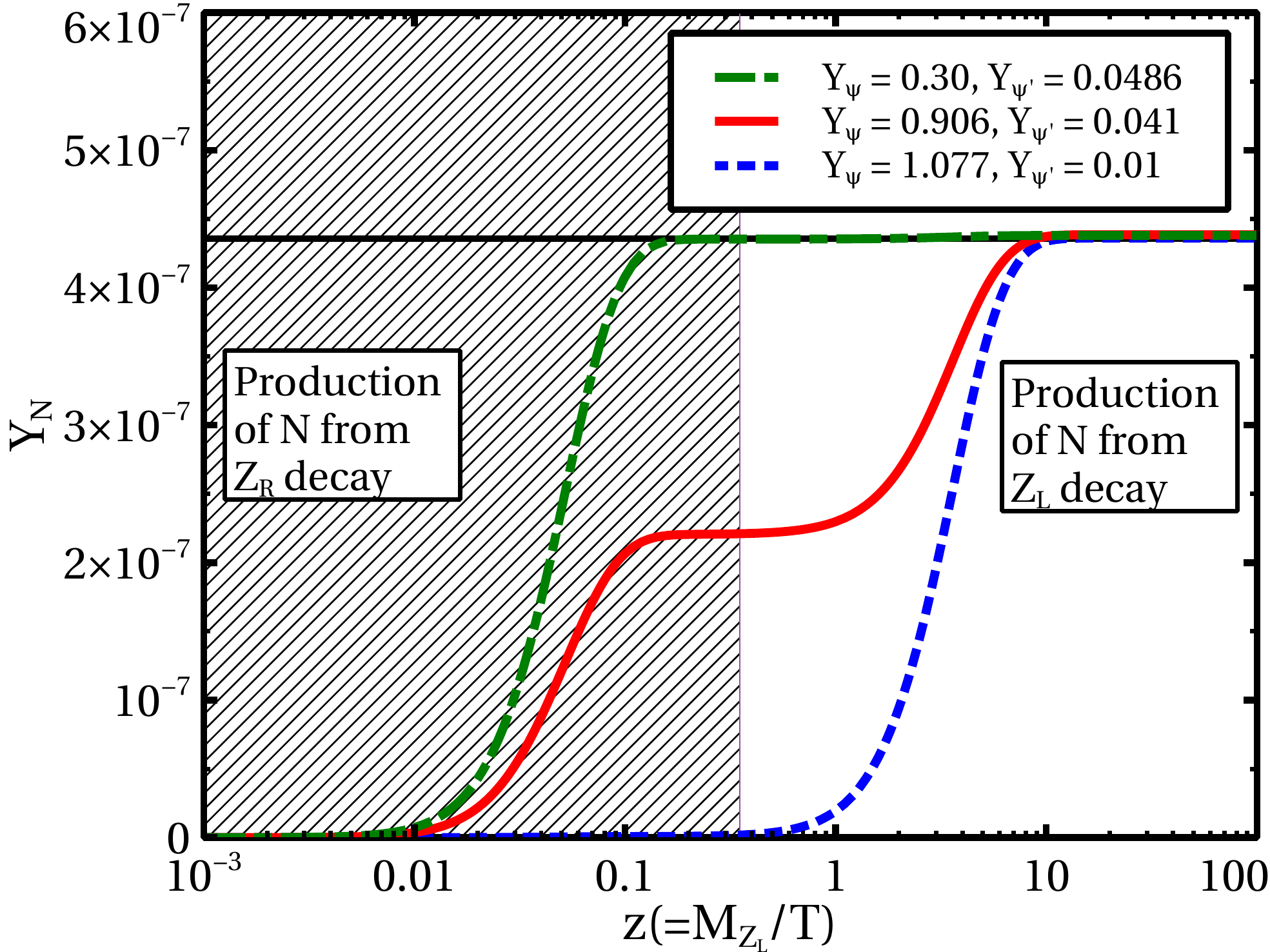}
\includegraphics[scale=0.4]{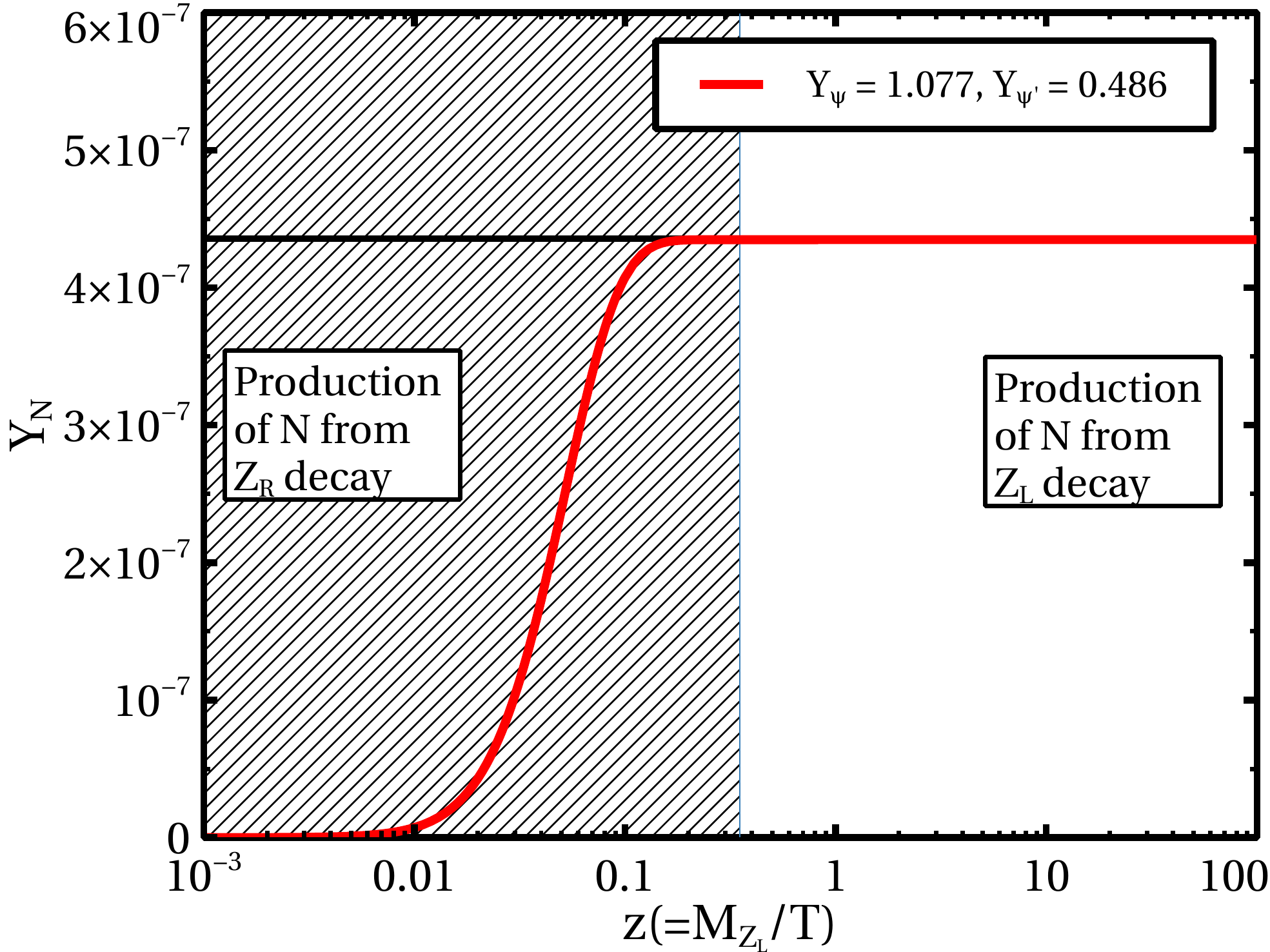}
\caption{Production of $N$ in the early Universe when
$z$ changes from 0.001 to 100 for $M_L=M_R=10^5$ GeV (left panel)
and $M_L=M_R=10^6$ GeV (right panel). For $M_L=M_R=10^5$ GeV, $N$
can be produced from both $Z_L$ and $Z_R$ depending upon the
respective Yukawa couplings, however for $M_L=M_R=10^6$ GeV,
the decay of $Z_R$ is the only dominant production mode of $N$ if we
restrict $Y_{\psi}$ within the perturbative limit. Black solid line in
both panels represents $Y_N = 4.356\times10^{-7}$, which reproduces
$\Omega_N h^2 = 0.12$ for $M_N=1$ MeV.}
\label{Fig:relic_density}
\end{figure}
Finally, the relic density of dark matter, which is defined as
the ratio between dark matter mass density and critical
density of the Universe, is obtained using the solution
of Boltzmann equation \footnote{In Appendix \ref{BEanalytical},
we have derived the analytical solution of the Boltzmann equation
involving only decay terms in the R.H.S.} at the present epoch in the following
equation \cite{Edsjo:1997bg}:
\begin{eqnarray}
\Omega_{\rm DM} h^2 = 2.755\times 10^8\left(\dfrac{M_N}
{\rm GeV}\right)\,Y_{N}(T_0)\,,
\label{Eq:omegaYrelation}
\end{eqnarray}  
where $T_0$ is the present temperature of
the Universe $\sim\mathcal{O}(10^{-13})$ GeV.
\subsection{Numerical results}
In both panels of Fig.\,\ref{Fig:relic_density}, we show the variation of
comoving number density ($Y_N$) of $N$ with the dimensionless variable
$z=\frac{M_{Z_L}}{T}$. This plot has been generated
for $M_{Z_R}=6$ TeV, $M_N=1$ MeV and three different
values of Yukawa couplings $Y_{\psi}$ and $Y_{\psi^{\prime}}$
which reproduce the correct dark matter relic density.
Moreover, in order to kinematically forbid
the tree level decay of either $H_L$ ($H_R$) or $\psi$ ($\psi^{\prime}$)
as well as to use the simple expressions for one loop decay widths in Eq.\,\,\eqref{loopfimp}
we choose $M_{H_L}=M_{\psi}=M_L$ ($M_{H_R}=M_{\psi^{\prime}}=M_R$). As mentioned earlier, we have also considered $M_L$, $M_R$, $M$, $M^{\prime}$
to be very large at least greater than the reheat temperature of the
Universe after inflation, which can be sufficiently low (but higher than the mass scale of the decaying particles) in some cosmological scenarios like for example, \cite{Moroi:1993mb, deGouvea:1997afu}. As a
result, the production of $N$ from the scatterings
of these heavy particles is not efficient.
The plot in the left panel is for $M_L=M_R=10^5$ GeV.
From this plot, it is seen that at first comoving number density
of $N$ increases as $z$ increases from $0.01$ to $0.1$
(corresponding temperature decreases from $M_{Z_L}/0.01$ GeV to $M_{Z_L}/0.1$ GeV) and
then for $0.1\leq z \leq0.35$,~$Y_N$
saturates to a particular value which depends upon the
value of Yukawa coupling $Y_{\psi^{\prime}}$ (e.g. 
green dashed-dotted line and red solid line). This initial rise in
the comoving number density of $N$ is
due to its production from the heavy right handed neutral gauge
boson $Z_R$. Since $Z_R$ is in equilibrium with the thermal
bath, most of the production of $N$ from the decay of $Z_R$
occurs for the temperature of the Universe $T\sim M_{Z_R}$. As the temperature
drops below the mass of $Z_R$, the number density of $Z_R$ starts
becoming exponentially suppressed (Boltzmann suppressed) and
finally for $z\geq 0.1$ (or $T\leq$ 1 TeV) there are practically
not enough number of right handed neutral gauge boson left
to produce $N$ and thus the comoving number density $Y_N$ saturates. Thereafter, again there
is a sharp increase in the comoving number density of $N$ between
$z=1.0$ to $z=10$ (e.g. red solid line and blue dotted line).
This increment of relic density is due to the substantial production
of $N$ from the decay of left handed neutral gauge boson $Z_L$
(the usual $Z$ boson in the SM), which depends on the other
Yukawa coupling $Y_{\psi}$  (see Eqs.\,\,\eqref{loopfimp} and
\eqref{gammaZLR}). Like the previous production
regime of $N$ from $Z_R$, in this case also the dominant production
of $N$ from $Z_L$ decay occurs at around the temperature
$T\sim M_{Z_L}$ ($z\sim 1$). Finally, when the temperature
of the Universe drops well below the mass of $Z_L$, all
the production modes of $N$ cease and $Y_N$ saturates
to $4.356\times10^{-7}$, the value of $Y_N$ which reproduces
$\Omega_{\rm DM}h^2=0.12$ for $M_N=1$ MeV. Here, we have chosen three
combinations of Yukawa couplings $Y_{\psi}$ and $Y_{\psi^{\prime}}$
which result in the correct relic density of dark matter. For $Y_{\psi}=0.906$
and $Y_{\psi^{\prime}}=0.041$, we have a situation where there is
an equal contribution of both $Z_R$ and $Z_L$ to $Y_N$
and that is represented by the red solid line in Fig.\,\,\ref{Fig:relic_density}.
On the other hand by tuning both $Y_{\psi}$ and $Y_{\psi^{\prime}}$,
one can have scenarios where the production of $N$ is dominated by
either $Z_R$ or $Z_L$. These are described by green dashed-dotted line
and blue dotted line respectively while the corresponding Yukawa
couplings are $Y_{\psi}=0.30$,
$Y_{\psi^{\prime}}=0.0486$
and $Y_{\psi}=1.077$, $Y_{\psi^{\prime}}=0.01$ respectively.
Similarly, in the right panel we have shown the variation of $Y_N$
with $z$ for $M_L=M_R=10^6$ GeV. However, unlike the plot in the left panel,
here our DM candidate $N$ is almost entirely produced from the decay of $Z_R$.
This can be understood if we notice the expression of loop factor in the
limit $M_{H_L}>>M_{Z_L}$, where the loop factor is proportional to $x=M^2_{Z_L}/M^2_{H_L}$.
Now, when we increase $M_{L}$ ($=M_{H_L}$) from $10^{5}$ GeV to $10^6$ GeV, the corresponding
loop factor for $Z_L$ becomes suppressed by a factor of $10^{2}$ and hence we need large
Yukawa coupling $Y_{\psi}$ to compensate this suppression. We have found that one
cannot get any significant contribution from $Z_L$ for $M_L=10^6$ GeV as
long as $Y_{\psi}$ remains within the perturbative limit ($Y_{\psi}\leq \sqrt{4\pi}$).

\begin{figure}
\includegraphics[scale=0.4]{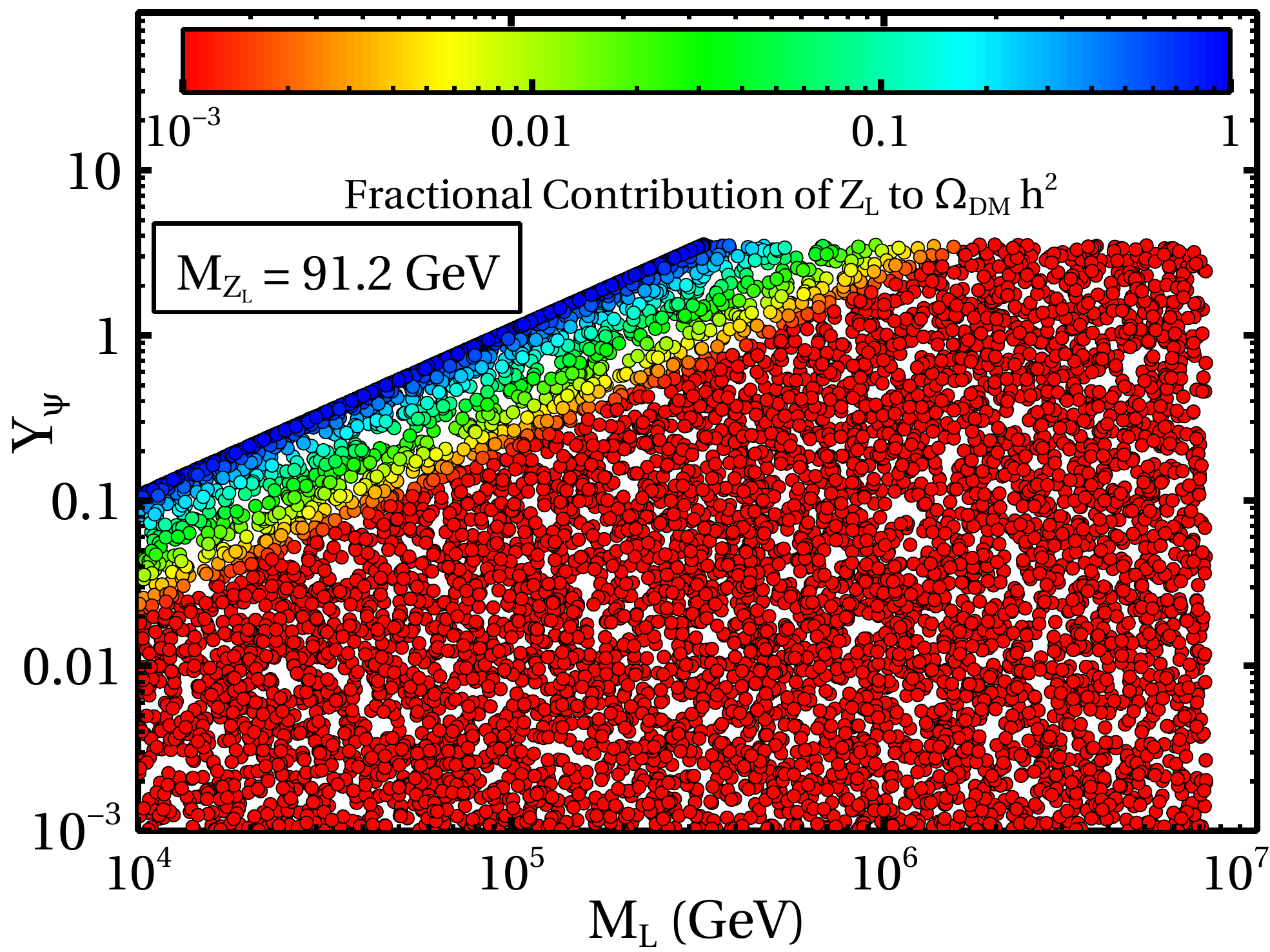}
\includegraphics[scale=0.4]{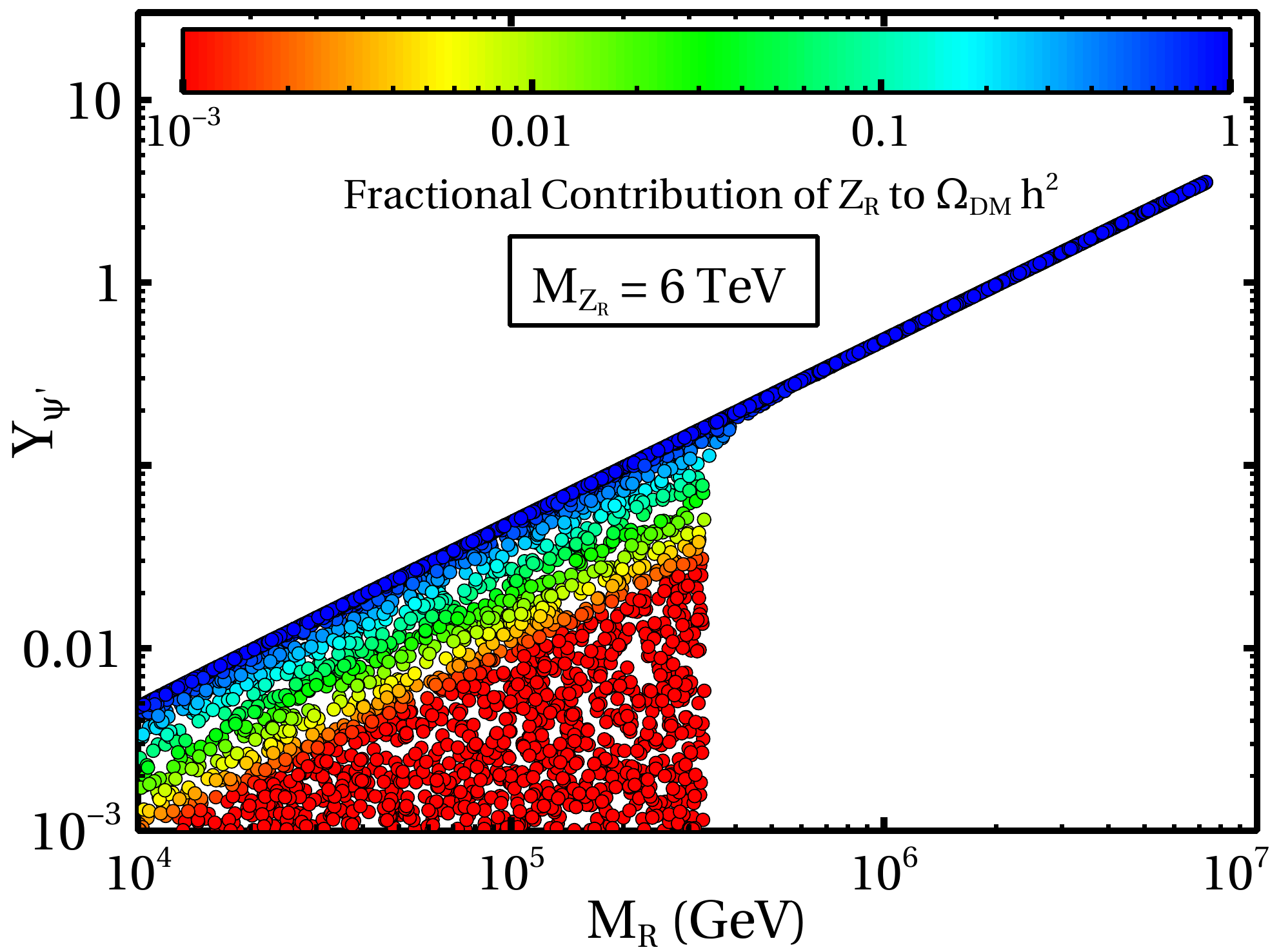}
\caption{Left panel: Variation of $Y_{\psi}$ with $M_L$. Right panel:
Variation of $Y_{\psi^{\prime}}$ with $M_R$. Colour code in each
panel represents the fractional contribution of the
respective gauge boson to $\Omega_{\rm DM}h^2$.}
\label{Fig:y-vs-M_0}
\end{figure}

The variation of Yukawa couplings $Y_{\psi}$ and $Y_{\psi^{\prime}}$
with $M_L$ and $M_R$ respectively has been shown in both panels of Fig.\,\,\ref{Fig:y-vs-M_0}.
In the left panel, we have shown the allowed parameter space in
$Y_{\psi}-M_L$ plane which reproduces the observed dark matter
relic density within $1\sigma$ limit. Now as we change $Y_{\psi}$
and $M_L$, the loop factor for $Z_L$ changes, which in turn
modifies the contribution of $Z_L$ to the relic density of $N$.
The variation of the fractional contribution of $Z_L$
(${\Omega_{\rm DM}^{Z_{L} \rightarrow \bar{N}N}}/{\Omega_{\rm DM}}$)
is shown by the colour code. The similar parameter space ($Y_{\psi^{\prime}}-M_R$)
for the right handed gauge boson $Z_R$ is also shown in the right panel of
Fig.\,\,\ref{Fig:y-vs-M_0}. From the plot in the right panel it is seen
that the fractional contribution of $Z_R$ to $\Omega_{\rm DM} h^2$
varies between $10^{-3}$ to 1 for $M_R\leq 4\times 10^{5}$ GeV.  
Thereafter, $Z_R$ becomes the dominant contributor  
for $4\times 10^{5}\,\,{\rm GeV}\,< M_R < 8\times 10^6$ GeV
and beyond that ($M_R\geq8\times 10^6$ GeV) there is no allowed
parameter space. This can be understood using the expression
of loop factor for $Z_R(Z_L)$ in the limit $M^2_{Z_R}(M^2_{Z_L})<<M_R^2(M^2_L)$.
In this limit, the loop factor for $Z_R(Z_L)$ is proportional to 
$\dfrac{M^2_R}{M^2_{Z_R}}\left(\dfrac{M^2_L}{M^2_{Z_L}}\right)$. Now, any increment in $M_R$
and $M_L$ decreases the corresponding loop factor and hence the contribution
of the respective gauge boson to $\Omega_{\rm DM} h^2$. This requires an
enhancement in the Yukawa couplings. However, since $M_{Z_L}<<M_{Z_R}$,
the suppression of loop factor for $Z_L$ is much more compared
to that of $Z_R$ for a particular value of $M_L$ and $M_R$. Therefore,
the contribution of $Z_L$ becomes negligibly small for $M_L = M_R \gtrsim 5\times 10^5$ GeV
and $Y_{\psi}\leq \sqrt{4\pi}$ where the entire $N$ production
occurs from the decay of $Z_R$ only. Similarly, if we keep on
increasing $M_R$ from $10^5$ GeV to $10^7$ GeV, we will encounter
a situation for $M_R > 8\times 10^6$ GeV when the loop factor
of $Z_R$ also becomes too small such that the Yukawa coupling
$Y_{\psi^{\prime}}$ within the perturbative limit is not enough
to produce the right DM abundance.

\begin{figure}
\includegraphics[scale=0.4]{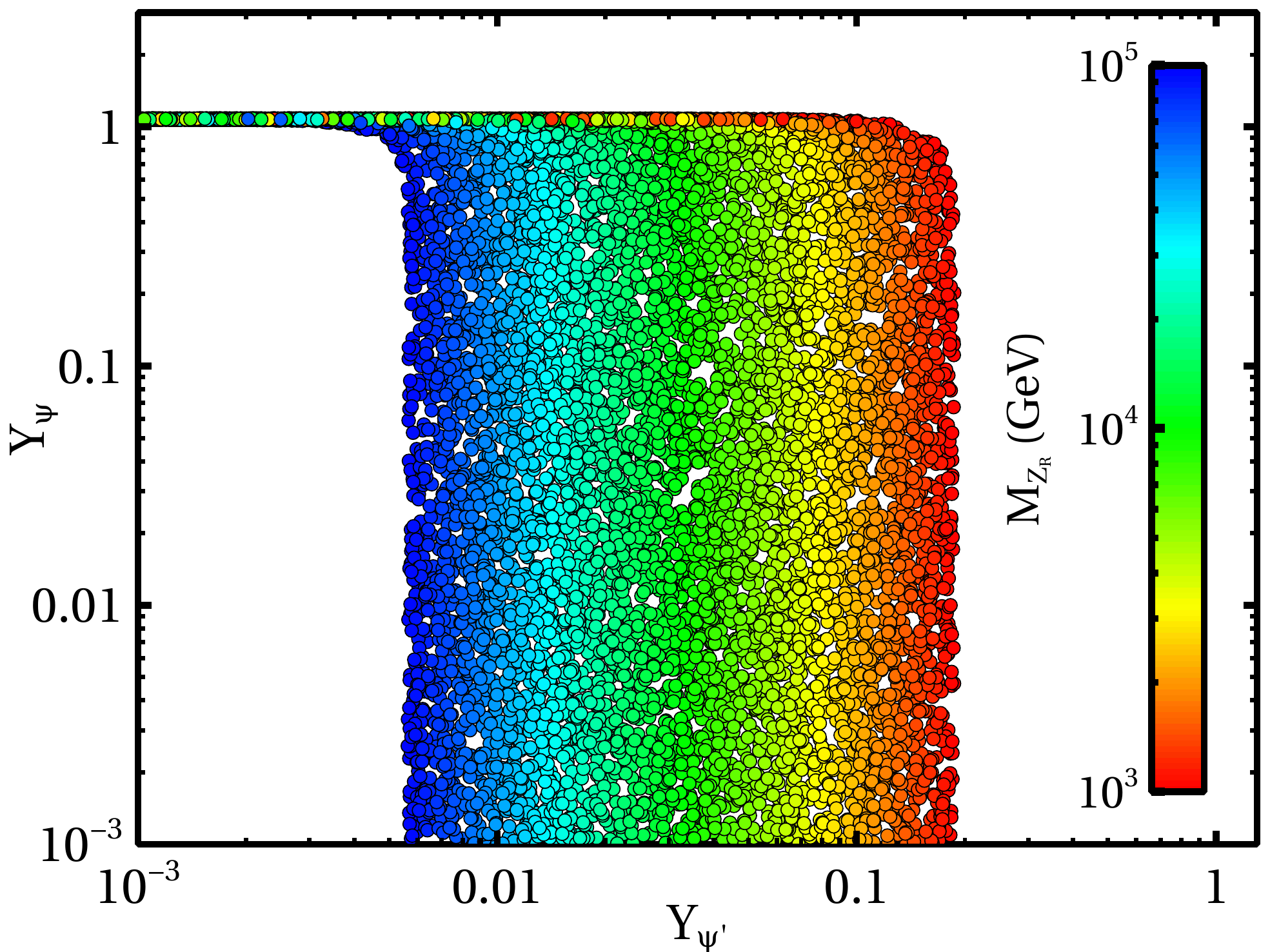}
\includegraphics[scale=0.4]{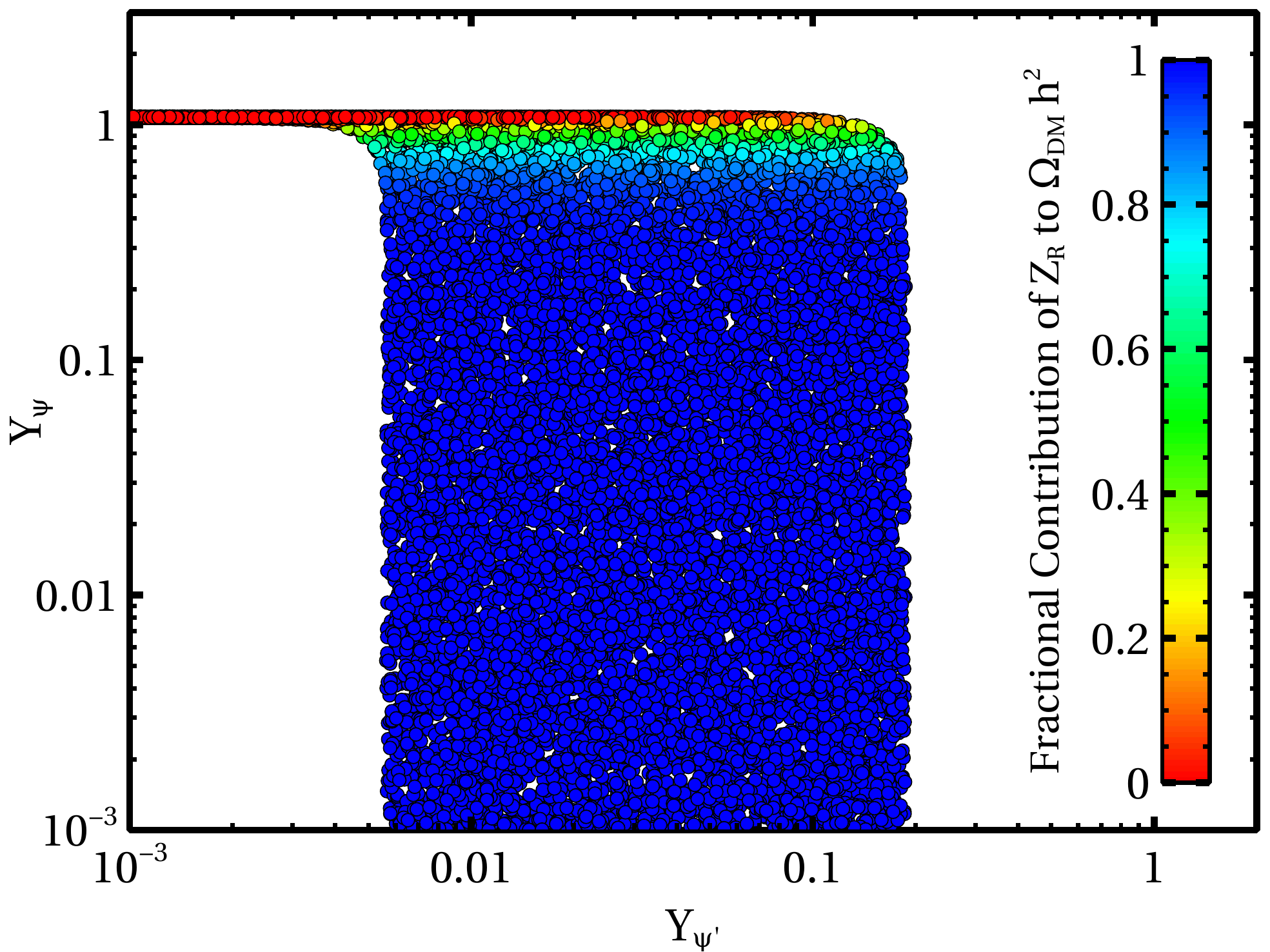}
\caption{Allowed region in $Y_{\psi}-Y_{\psi^{\prime}}$ plane
which reproduces correct dark matter relic density for
$1\,{\rm TeV}\,\leq M_{Z_R}\leq\,100$ TeV. We also
shown the variation of $M_{Z_R}$ (left panel) and the variation
of the fractional contribution of $Z_R$ to the relic density of $N$ (right panel).}
\label{Fig:ypsi-ypsip}
\end{figure} 

In both panels of Fig.\,\,\ref{Fig:ypsi-ypsip}, we demonstrate the region
in $Y_{\psi}-Y_{\psi^{\prime}}$ plane which is allowed
by the observed value of dark matter relic density at 68\%
C.L. ($0.1172 \leq \Omega_{\rm DM} h^2 \leq 0.1226$).
While generating these two plots we have scanned
over the Yukawa couplings $Y_{\psi}$ and $Y_{\psi^{\prime}}$
between $10^{-3}\leq Y_{\psi^{(\prime)}} \leq \sqrt{4\pi}$
and the mass of ${Z_R}$ in the range of 1 TeV to 100 TeV.
The corresponding variation of $M_{Z_R}$ in $Y_{\psi^\prime}-Y_{\psi}$
plane has been shown by the colour code in the left panel.
From this plot, one can notice that we need larger values of
Yukawa coupling $Y_{\psi^\prime}$ when $M_{Z_R}$ decreases
from $10^{5}$ GeV to $10^{3}$ GeV. This is because, the loop factor
in the limit $M^2_{Z_R}<<M^2_R$ decreases with $x=M^2_{Z_R}/M^2_R$
and we need to enhance the Yukawa coupling $Y_{\psi^\prime}$ appropriately
to bring back the same contribution of $Z_R$ to $\Omega_{\rm DM} h^2$. 
In the right panel, we have drawn the same parameter
space but in this case the colour code is used
to demonstrate how the fractional contribution of $Z_R$ to $\Omega_{\rm DM} h^2$
changes with respect to the variation of the Yukawa couplings. The
information about the fractional contribution of $Z_L$ can also be obtained
from this plot by subtracting the contribution of $Z_R$ 
from unity.\,\,From this plot it is clearly seen that
for $Y_{\psi^{\prime}}>5\times 10^{-3}$ and $Y_{\psi}<0.7$,
the production of $N$ is dominated by the $Z_R$ decay
and on the other hand, the relic density of $N$ receives
maximum contribution from $Z_L$ when $Y_{\psi}\sim 1$.
In the narrow intermediate region ($0.6\leq Y_{\psi}\leq 1.0$
and $6\times 10^{-3}\leq Y_{\psi^{\prime}}\leq 0.2$), both $Z_L$ and
$Z_R$ are contributing to the relic density
of $N$ and their contributions depend upon the specific
values of $Y_{\psi}$ and $Y_{\psi^{\prime}}$.
Moreover, we get a narrow horizontal line for $Y_{\psi}\sim 1$
and $Y_{\psi^{\prime}}\lesssim 6\times10^{-3}$,
where the entire production of $N$ occurs from $Z_L$ decay. The opposite
situation where the entire $N$ is produced from $Z_R$ decay
occurs for $Y_{\psi}< 0.6$ and $Y_{\psi^{\prime}}> 6\times10^{-3}$.
However, unlike to the $Z_L$ dominated case, here we get a wide
band and it due to the variation of mass of $Z_R$,
which has been varied between 1 TeV to 100 TeV.
\begin{figure}
\includegraphics[scale=0.55]{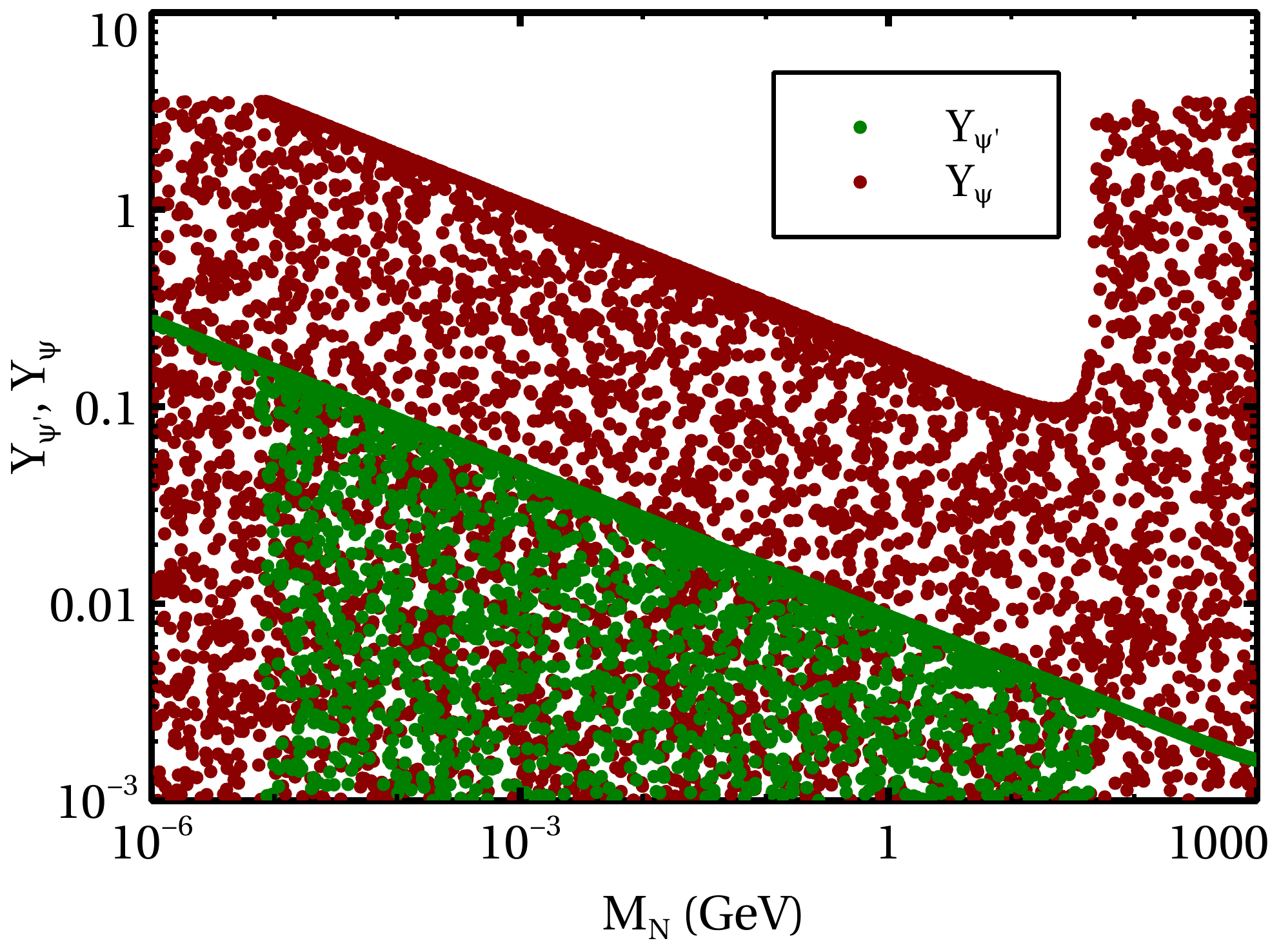}
\caption{The allowed values $Y_{\psi}$ (dark red coloured contour)
and $Y_{\psi^{\prime}}$ (green coloured contour) from correct relic density
criteria when $M_N$ varies from 1 keV to 1 TeV.}
\label{Fig:mn-vs-y}
\end{figure} 

Finally, we have also shown the variation of Yukawa couplings
$Y_{\psi}$ and $Y_{\psi^{\prime}}$ with the mass of our dark
matter candidate $N$ in Fig.\,\,\ref{Fig:mn-vs-y}. In this
plot, we have varied $M_N$ in the range of 1 keV to 1 TeV. The
corresponding variation of $Y_{\psi^{\prime}}$ ($Y_{\psi}$) to
obtain the correct dark matter relic abundance is indicated by
green (dark red) coloured contour. From this plot, one can notice
that the allowed values of Yukawa couplings are decreasing with
the increase of $M_N$. This can be understood from
Eq.\,\,(\ref{Eq:omegaYrelation}) which states that for a
fixed value of $\Omega_{\rm DM}h^2$ the product of
$Y_{N}(t_0)$ and $M_N$ is a constant. Since, our
dark matter candidate is a FIMP, $Y_{N}(t_0)$
is proportional to the Yukawa couplings. Therefore,
to remain within the observed relic density band, any increment
in $M_N$ must be accompanied by a decrement in $Y_{N}(t_0)$
and consequently in $Y_{\psi}$ and $Y_{\psi^{\prime}}$. In the
right most corner of $M_N-Y_{\psi}$ plane, one can see that
the entire range of $Y_{\psi}$ is allowed for $M_N>M_{Z_L}/2$.
This is due to the fact that in this mass range of $N$, the
production from $Z_L$ decay is kinematically forbidden and
hence any variation in $Y_{\psi}$ does not affect the
the relic density of $N$. On the other hand, as $Z_L$ decay to a pair
of $N$ is not possible for $M_N>M_{Z_L}/2$, the decay
$Z_R$ is solely responsible for the entire production
of $N$. Therefore, in this mass range of $M_N$,
we get a very narrow allowed values of $Y_{\psi^{\prime}}$.
On the other hand, for low mass DM ($M_N\leq 5 \times 10^{-6}$ GeV)
we also get a narrow band for $Y_{\psi^{\prime}}$ while
all the values of other Yukawa coupling $Y_{\psi}$
are allowed. This due to the fact that for the low
dark matter mass, we need large $Y_N(t_0)$ to satisfy
the relic density criterion as $\Omega_{DM} h^2$
is directly proportional to $M_N$ and for $Z_L$ 
to contribute significantly to $Y_N(t_0)$ we need
$Y_{\psi}$ beyond the perturbative regime. In other
words in this low mass DM region $Z_R$ again 
becomes the dominant contributor to DM relic
abundance. Nevertheless, as the entire considered mass range of $N$
is allowed for some combinations of $Y_{\psi}$ and $Y_{\psi^{\prime}}$,
we have checked the nature of our dark matter candidate (hot/warm/cold)
by computing its free streaming length following Ref. \cite{Merle:2013wta}. 
We find that $N$ becomes a warm dark matter
candidate for $M_N\leq 10$ keV, where its
free streaming length goes above $0.01\,\,{\rm Mpc}$. 
On the other hand, the cold dark matter scenario
is viable for $M_N> 10$ keV, where the free streaming
length of $N$ always lies below 0.01 Mpc and
decreases sharply with the increase
of mass of $N$. The possibility of warm dark matter has several motivations,
which can be found in the recent review \cite{Adhikari:2016bei}.
\section{Conclusion}
\label{sec:conc}
We have proposed a UV complete framework to dynamically generate tiny couplings required for non-thermal dark matter scenarios whose relic abundance is generated through the freeze-in mechanism, within the framework called freeze-in massive particle. Based on gauge symmetric extensions of the Standard Model, we particularly consider the left-right symmetric model which have several other motivations related to the origin of parity violation, neutrino mass among others. Considering the dark matter candidate to be a gauge singlet fermion which has no tree level couplings with the Standard Model particles, we generate its couplings with the Standard Model particles at one loop, mediated by gauge bosons and Higgs. After showing such a dark matter candidate to remain out of thermal equilibrium in the early Universe for generic choices Yukawa couplings and masses of particles inside loops, we then calculate its relic abundance by considering both decay and scattering contributions in a way similar to a generic FIMP dark matter candidate. We find that the decay of neutral heavy gauge bosons to a pair of FIMP dark matter candidate is the most dominant production mechanism and can give rise to the correct relic abundance for Yukawa couplings as large as 
$\mathcal{O}(0.01)$ to $\mathcal{O}(1)$ while keeping the additional heavy neutral boson mass within experimental reach. Such Yukawa couplings lie in a range which less fine tuned than the electron Yukawa coupling in the SM and far less fine tuned than the ones involved in generic FIMP models. On the other hand, a very wide range of dark matter masses is consistent with the relic abundance criteria and some portion of this allowed range can also give rise to warm dark matter scenarios that have several other motivations from small scale structure point of view. Since our UV complete setup has many other particles that lie in the experimentally accessible range, many associated particles can be probed at ongoing experiments, which we leave for future studies. We also note that such a setup can be realised in other gauge extensions of Standard Model as well which may be relatively simpler than the one presented here as a matter of choice.

\acknowledgments
DB acknowledges the support from IIT Guwahati
start-up grant (reference number: xPHYSUGIITG01152xxDB001)
and Associateship Programme of IUCAA, Pune. One of the authors
AB acknowledges the financial support from SERB, Govt. of INDIA
through NPDF fellowship with grant no. PDF/2017/000490. AB also
gratefully acknowledges the cluster computing facility at
HRI, Allahabad (http://cluster.hri.res.in).
\appendix
\section{Scalar Potential of the Model}
\label{appen1}
The scalar potential for the minimal LRSM is 
\begin{equation}
 V(\Phi,\Delta_L,\Delta_R) = V_{\mu} + V_{\Phi} + V_{\Delta} + V_{\Phi\Delta} + V_{\Phi\Delta_L\Delta_R},
\end{equation}
where the bilinear terms in Higgs fields are
\begin{eqnarray}
 V_{\mu} &=& -\mu_1^2 \Tr{\Phi^\dagger\Phi}
 - \mu_2^2\Tr{\Phi^\dagger\tilde{\Phi} + \tilde{\Phi}^\dagger\Phi}
 - \mu_3^2\Tr{\Delta_L^\dagger\Delta_L + \Delta_R^\dagger\Delta_R}.
 \end{eqnarray}
The self-interaction terms of $\Phi$ are: 
\begin{align}
 V_{\Phi} = ~
  & \lambda_1\left[\Tr{\Phi^\dagger\Phi}\right]^2 +
  \lambda_2\left[\Tr{\Phi^\dagger\tilde{\Phi}}\right]^2 + \lambda_2\left[\Tr{\tilde{\Phi}^\dagger\Phi}\right]^2 
  \nonumber \\   
  & + \lambda_3\Tr{\Phi^\dagger\tilde{\Phi}}\Tr{\tilde{\Phi}^\dagger\Phi} + 
  \lambda_4\Tr{\Phi^\dagger\Phi}\Tr{\Phi^\dagger\tilde{\Phi} + \tilde{\Phi}^\dagger\Phi}.
  \label{appeneq1}
\end{align}
and the $\Delta_{L,R}$ self- and cross-couplings are as follows:
\begin{align}
 V_{\Delta} = ~ 
 & \rho_1\left(\left[\Tr{\Delta_L^\dagger\Delta_L}\right]^2 + 
     \left[\Tr{\Delta_R^\dagger\Delta_R}\right]^2\right) +
 \rho_3\Tr{\Delta_L^\dagger\Delta_L}\Tr{\Delta_R^\dagger\Delta_R}
 \nonumber \\ 
 & + \rho_2 \left(\Tr{\Delta_L\Delta_L}\Tr{\Delta_L^\dagger\Delta_L^\dagger} + \Tr{\Delta_R\Delta_R}\Tr{\Delta_R^\dagger\Delta_R^\dagger} \right)
\nonumber  \\ 
 & + \rho_4\left(\Tr{\Delta_L\Delta_L}\Tr{\Delta_R^\dagger\Delta_R^\dagger} + \Tr{\Delta_L^\dagger\Delta_L^\dagger}\Tr{\Delta_R\Delta_R}\right).
   \label{appeneq2}
\end{align}
In addition, there are also  $\Phi-\Delta_L$ and $\Phi-\Delta_R$ interactions present in the model, 
\begin{align}
 V_{\Phi\Delta} = ~
 & \alpha_1\Tr{\Phi^\dagger\Phi}\Tr{\Delta_L^\dagger\Delta_L + \Delta_R^\dagger\Delta_R} +
 \alpha_3 \Tr{\Phi\Phi^\dagger\Delta_L\Delta_L^\dagger + \Phi^\dagger\Phi\Delta_R\Delta_R^\dagger} 
  \nonumber \\ 
 & + \left\{
 \alpha_2 e^{i\delta_2}\Tr{\Phi^\dagger\tilde{\Phi}}\Tr{\Delta_L^\dagger\Delta_L} + 
 \alpha_2 e^{i\delta_2}\Tr{\tilde{\Phi}^\dagger\Phi}\Tr{\Delta_R^\dagger\Delta_R} + \text{H.c.}\right\}
  \label{appeneq3}
\end{align}
with $\delta_2 =0$ making CP conservation explicit, and the $\Phi-\Delta_L-\Delta_R$ couplings are
\begin{align}
 V_{\Phi\Delta_L\Delta_R} = ~
 & \beta_1\Tr{\Phi^\dagger\Delta_L^\dagger\Phi\Delta_R + \Delta_R^\dagger\Phi^\dagger\Delta_L\Phi} +
 \beta_2\Tr{\Phi^\dagger\Delta_L^\dagger\tilde{\Phi}\Delta_R + \Delta_R^\dagger\tilde{\Phi}^\dagger\Delta_L\Phi}
\nonumber \\ 
& + \beta_3\Tr{\tilde{\Phi}^\dagger\Delta_L^\dagger\Phi\Delta_R + \Delta_R^\dagger\Phi^\dagger\Delta_L\tilde{\Phi}}.
\end{align}
The scalar potential involving the newly introduced scalar fields beyond the minimal LRSM is
\begin{equation}
V_{\rm new} = V_{H} + V_{\sigma} + V_{\Phi H}+ V_{\Delta H}.
\end{equation}
The details of different terms on the right-hand side of the above equation can be written as follows,
\begin{align}
V_{H} = ~ & \mu^2_{H} (H_L^{\dagger} H_L +H_R^{\dagger} H_R) + \rho_5 (\left[H_L^{\dagger} H_L\right]^2 +\left[H_R^{\dagger} H_R\right]^2) \nonumber \\
& +\rho_6 \left[H_L^{\dagger} H_L\right]\left[H_R^{\dagger} H_R\right],
\end{align}
\begin{align}
V_{\sigma}  = ~ & \frac{\mu^2_{\sigma}}{2} \sigma^2 + \rho_8  \sigma^4 + \mu_{\sigma \Delta} \sigma (\Tr{\Delta_L^\dagger\Delta_L - \Delta_R^\dagger\Delta_R}) +\mu_{\sigma H} \sigma (H_L^{\dagger} H_L-H_R^{\dagger} H_R)  \nonumber \\
& +\rho_9 \sigma^2 \Tr{\Phi^\dagger\Phi}+\rho_{10} \sigma^2 (\Tr{\Delta_R^\dagger\Delta_R + \Delta_L^\dagger\Delta_L}) + \rho_{11} \sigma^2 (H_L^{\dagger} H_L+H_R^{\dagger} H_R),\end{align}
\begin{align}
 V_{\Phi H}  = ~
& \mu_{14}H^\dagger_L \Phi H_R+f_{145} \Tr{\Phi^{\dagger} \Phi} (H_L^{\dagger} H_L+H_R^{\dagger} H_R),
 \end{align}
\begin{align}
 V_{\Delta H}  = ~
& (\mu_{15}H_L \Delta_L H_L+ \mu_{16}H_R \Delta_R H_R+\text{H.c.})+f_{145} (\Tr{\Delta^{\dagger}_L \Delta_L}+\Tr{\Delta^{\dagger}_R \Delta_R}) (H_L^{\dagger} H_L+H_R^{\dagger} H_R) 
 \end{align}
\section{Decay width calculation}
\label{loopcalc} 
In order to calculate the ``Feeble'' interaction which shows up radiatively,
we would first like to briefly discuss the Lagrangian from which different vertices in the
loop arise. The vertex involving gauge bosons and fermion/scalars will arise from the respective
kinetic terms involving covariant derivatives. The covariant derivative for the gauge group of
Left-Right model can be written as:
\begin{align}
    D^\mu_{L,R} &= \left(\partial_\mu - i g_{L,R} \frac{\vec{\tau}}{2}\vec{W}_{L,R} - ig_{B-L}\frac{({\bf B-L})}{2}B_\mu\right)
\end{align}
Now, from the above covariant derivative the kinetic part of the Lagrangian for fermion doublets $\psi$, $\psi'$ and scalar doublets $H_L, H_R$ are as follows:
\begin{align}
    \mathcal{L}_{\rm kin} \subset i\overline{\psi}\gamma_\mu D^\mu_L \psi + i \overline{\psi'}\gamma_\mu D^\mu_R \psi' + (D_{\mu R} H_R)^\dagger(D^\mu_R H_R) + (D_{\mu L} H_L)^{\dagger}(D^\mu_L H_L)
\end{align}
On the other hand, the scalar-fermion-fermion vertices in the loop diagrams arise from the corresponding Yukawa interactions shown in \eqref{yukawa1}. Thus, the relevant interaction terms contributing to the one loop decay width 
of $Z_{L,R}$ into a pair of $N$'s are 
\begin{align}
    \mathcal{L}_{\rm feeble} &\subset \Tilde{g}_{L}\overline{\psi}\gamma_\mu\psi Z^\mu_L + \Tilde{g}_{R}\overline{\psi'}\gamma_\mu\psi' Z^\mu_R 
    - i\Tilde{g}_L (\partial_\mu H^\dagger_L H_L - H^\dagger_L \partial_\mu H_L)Z^\mu_L \nonumber \\ 
    &- i\Tilde{g}_R (\partial_\mu H^\dagger_R H_R - H^\dagger_R \partial_\mu H_R)Z^\mu_R  + Y_\psi \overline{\psi}\Tilde{H}_L N + Y_{\psi'} \overline{\psi'}\Tilde{H}_R N
\end{align}
where $\Tilde{g}_{L, R}$ are the respective gauge couplings in the physical basis of the neutral gauge bosons. The decay of $Z_{L,R}$ to pair of FIMP's ($N$'s) from the above Lagrangian are 
shown in Fig.\ref{fig:Z_decay} and we closely follow the two-spinor technique \cite{Dreiner:2008tw} in 
order to calculate the loop factors and the aforementioned figure is also shown using the same notations.

The Lagrangian for the above decay process is $-i\mathcal{A}\overline{N}\gamma^\mu\gamma^5 Z_{\mu L, R} N$ where the loop factor $\mathcal{A}$ is given as:
\begin{align}
   \mathcal{A} &= \frac{i\Tilde{g}Y^2}{16\pi^2}\left[1 + \frac{m^2_{H_{L,R}}}{M^2_{Z_{L,R}}}\ln\left[\frac{2m^2_{H_{L,R}}-M^2_{Z_{L,R}} + \sqrt{M^4_{Z_{L,R}} - 4 m^2_{H_{L,R}}M^2_{Z_{L,R}}}}{2m^2_{H_{L,R}}}\right]^2\right]. 
\end{align}
If we take the limit $m^2_{H_{L,R}}\ll M^2_{Z_{L,R}}$ and use the parametrisation $y=\frac{m^2_{H_{L,R}}}{M^2_{Z_{L,R}}}$, then the loop factor is
\begin{align*}
    \mathcal{A} &= \frac{i\Tilde{g}Y^2}{16\pi^2}\left[1 + y\ln[y]^2 - \pi^2y + 2i\pi \ln y + \mathcal{O}(y^2)\right].
\end{align*}
On the other hand, going to the other limit $m^2_{H_{L,R}}\gg M^2_{Z_{L,R}}$ and parametrising $x=\frac{M^2_{Z_{L,R}}}{m^2_{H_{L,R}}}$ we can write the loop factor as
\begin{align*}
    \mathcal{A} &= \frac{i\Tilde{g}Y^2}{16\pi^2}\left[-\frac{x}{12} + \mathcal{O}(x^{3/2})\right]\,.
\end{align*}

It should be noted that, we have explicitly written the interaction term of gauge bosons with a pair of $N$'s as axial vector couplings. This is due to the Majorana nature of fermion $N$. Let us take a detour in showing why Majorana particle cannot have vector like interaction. Let us start with the following general interaction for Majorana particle $N=N^c$ (where $N^c$ is charge conjugated field of $N$)
\begin{align}
    \frac{1}{2}\overline{N}(g_V\gamma^\mu + g_A\gamma^\mu\gamma^5)N &=
    \frac{1}{2}\overline{N^c}(g_V\gamma^\mu + g_A\gamma^\mu \gamma^5) N^c \nonumber \\ 
    &= \frac{g_V}{2} \overline{C\overline{N}^T}\gamma^\mu C \overline{N}^T 
    + \frac{g_A}{2} \overline{C\overline{N}^T}\gamma^\mu\gamma^5 C \overline{N}^T \nonumber \\
    &= \frac{g_V}{2}(C\overline{N}^T)^\dagger\gamma^0 \gamma^\mu C \overline{N}^T 
    + \frac{g_A}{2}(C\overline{N}^T)^\dagger\gamma^0 \gamma^\mu CC^{-1} \gamma^5 C \overline{N}^T     \nonumber \\
    &= \frac{g_V}{2}{{\overline{N}^{}}^*}C^{-1}\gamma^0 \gamma^\mu C \overline{N}^T
    + \frac{g_A}{2}{{\overline{N}^{}}^*}C^{-1}\gamma^0 \gamma^\mu C {\gamma^{5}}^T \overline{N}^T \nonumber \\
    &= \frac{g_V}{2}N^T {\gamma^0}^* C^{-1}\gamma^0 C C^{-1}\gamma^\mu C\overline{N}^T 
    + \frac{g_A}{2}N^T {\gamma^0}^* C^{-1}\gamma^0 C C^{-1}\gamma^\mu C {\gamma^{5}}^T\overline{N}^T \nonumber \\
    &= -\frac{g_V}{2}N^T {\gamma^{0}}^*{\gamma^0}^T C^{-1}\gamma^\mu C \overline{N}^T -
     \frac{g_A}{2}N^T {\gamma^{0}}^*{\gamma^0}^T C^{-1} \gamma^\mu C {\gamma^{5}}^T \overline{N}^T \nonumber \\
    &= -\frac{g_V}{2}N^T C^{-1}\gamma^\mu C \overline{N}^T -
    \frac{g_A}{2} N^T C^{-1}\gamma^\mu C {\gamma^{5}}^T\overline{N}^T \nonumber \\
    &= + \frac{g_V}{2} N^T {{{\gamma^{\mu}}^{}}^T} \overline{N}^T + 
    \frac{g_A}{2} N^T {{{\gamma^{\mu}}^{}}^T} {\gamma^{5}}^T\overline{N}^T\nonumber \\
    &=  -\frac{g_V}{2}  \overline{N}\gamma^\mu N + 
    \frac{g_A}{2}  \overline{N}\gamma^\mu \gamma^5 N \nonumber \\
   g_V\overline{N}\gamma^\mu N &= 0
\end{align}
which justifies the use of axial vector like couplings of $N$'s with gauge bosons.
\section{Analytical solution of Boltzmann equation}
\label{BEanalytical} 
In this section, we present analytical solution of the Boltzmann
equation considering dark matter production through decays of heavy
particles only. Once we derive the analytical expression of comoving
number density ($Y_N$) for decays then the same procedure can be followed
to find $Y_N$ due to scattering processes. In terms of $Y_N$ the
Boltzmann equation for $N$, due to its production from the decays of
$Z_L$, $Z_R$ and $\Phi$, is given by,
\begin{eqnarray}
\dfrac{dY_{N}}{dz} &=& \dfrac{2 M_{\rm Pl}}{1.66 M_{sc}^{2}}
\dfrac{z \sqrt{g_{\star}(z)}}{g_{s}(z)}\,\,\Bigg[\sum_{\chi\,=\,Z_{L},
\,Z_{R},\,\Phi}\langle\Gamma_{\chi\rightarrow  \bar{N} N}
\rangle(Y_{\chi}^{\rm eq} - Y_{N})\Bigg]\,\,,
\label{BEdecay}
\end{eqnarray} 
where, $M_{sc}$ is some arbitrary mass scale which we consider equal
to $M_{Z_L}$. Now, since $N$ is a dark matter candidate of non-thermal
origin, we can neglect $Y_N$ in the R.H.S. of the above equation
due to its smallness compared to $Y_\chi$ ($\chi=Z_L,\,Z_R$ and $\Phi$)
of mother particles which are in thermal equilibrium and obey
the Maxwell-Boltzmann distribution. Hence $Y^{\rm eq}_{\chi}$ can be
expressed as
\begin{eqnarray}
Y^{\rm eq}_{\chi}(z) = \frac{45\,g_{\chi}}{4\pi^4}\,
\dfrac{r^2_{\chi}\,z^2\,{\rm K}_2(r_{\chi}\,z)}
{g_{s}(z)}\,,
\label{YNeq}
\end{eqnarray}
where, $r_{\chi}=\dfrac{M_{\chi}}{M_{sc}}$ is the ratio between
$M_{\chi}$ and $M_{sc}$. Moreover, $g_s(z)$ being the 
total degrees of freedom of all relativistic
species contributing to the entropy density
of the Universe at $z=M_{sc}/T$. The expression of
$\langle \Gamma_{\chi \rightarrow \bar{N} N}\rangle$ is given by
\begin{eqnarray}
\langle \Gamma_{\chi \,\rightarrow\bar{N}N} \rangle
&=& \Gamma_{\chi \,\rightarrow\bar{N}N}
\dfrac{{\rm K}_1\left(r_{\chi}\,z\right)}
{{\rm K}_2\left(r_{\chi}\,z\right)}\,.
\label{therAvrdeacy}
\end{eqnarray}
Substituting Eqs.\,\,(\ref{YNeq}) and (\ref{therAvrdeacy}) in
Eq.\,\,(\ref{BEdecay}) we get, 
\begin{eqnarray}
\dfrac{dY_{N}}{dz} &=& \dfrac{2\,M_{\rm Pl}}{1.66\,M_{sc}^{2}}\,
\frac{45}{4\pi^4}\,\dfrac{\sqrt{g_{\star}(z)}}{g_{s}(z)^2}
\sum_{\chi=Z_L,\,Z_R,\,\Phi} g_{\chi}\,r^2_{\chi}\,
\Gamma_{\chi\rightarrow\bar{N}N}\,{\rm K}_{1}(r_{\chi}\,z)\,z^3\,.
\label{BEdecay_step2}
\end{eqnarray}  
Now, $\dfrac{\sqrt{g_{\star}(z)}}{g_{s}(z)^2}\simeq
\dfrac{1}{\sqrt{g_{\rho}(z)}\,g_s(z)}$ where $g_{\rho}(z)$
is the total relativistic degrees of freedom contributing to the
energy density of the Universe at $z = M_{sc}/T$. Here, we have
neglected the a term proportional to $\dfrac{{\rm d}\,{\rm ln}
\,g_{\rm s}(z)}{{\rm d} \,{\rm ln} z}$ (see Eq.\,(\ref{gstarz})),
which becomes important only around the QCD phase transition temperature
($T\sim \mathcal{O}(100)$ MeV), where $g_s(z)$ changes drastically
with $z$ \cite{Gondolo:1990dk}. Therefore, the Eq.\,(\ref{BEdecay_step2})
simplifies to 
\begin{eqnarray}
\dfrac{dY_{N}}{dz} &=& \dfrac{2\,M_{\rm Pl}}{1.66}\,
\frac{45}{4\pi^4}\,\dfrac{1}{\sqrt{g_{\rho(z)}}\,g_s(z)}
\sum_{\chi=Z_L,\,Z_R,\,\Phi} 
\dfrac{g_{\chi}\,r^4_{\chi}}{M^2_{\chi}}\,\Gamma_{\chi\rightarrow\bar{N}N}
\,{\rm K}_{1}(r_{\chi}\,z)\,z^3\,,
\end{eqnarray}
and therefore
\begin{eqnarray}
Y_{N} = \dfrac{2\,M_{\rm Pl}}{1.66}\,
\frac{45}{4\pi^4}
\sum_{\chi=Z_L,\,Z_R,\,\Phi} 
\dfrac{g_{\chi}\,r^4_{\chi}}{M^2_{\chi}}\,\Gamma_{\chi\rightarrow\bar{N}N}
\,\int^{z_{max}}_{z_{min}}\dfrac{{\rm K}_{1}(r_{\chi}\,z)\,z^3}
{\sqrt{g_{\rho}(z)}\,g_s(z)}\,dz\,.
\end{eqnarray}
$z_{\{min,\,max\}}$ correspond to initial and final temperatures respectively.
One can further simplify the above equation to
\begin{eqnarray}
Y_{N} \simeq \dfrac{2\,M_{\rm Pl}}{1.66}\,
\frac{45}{4\pi^4}
\sum_{\chi=Z_L,\,Z_R,\,\Phi} 
\dfrac{g_{\chi}}{M^2_{\chi}\sqrt{g_{\rho}}\,g_s}\,\Gamma_{\chi\rightarrow\bar{N}N}
\,\int^{x_{max}}_{x_{min}}{\rm K}_{1}(x)\,x^3\,dx\,,
\end{eqnarray}
where $g_{\rho}$, $g_s$ are evaluated at $T=M_{\chi}$ and $x=r_{\chi}\,z$.
The above integral is maximum around $x \sim 1 $ (or $T\sim M_{\chi}$),
where we assume $g_{\rho}$, $g_s$ remain nearly constant. This is a reasonable
assumption as long as $T$ is far away from the QCD phase transition temperature.
Now, considering $x_{min}$ and $x_{max}$ equal to 0 and $\infty$,
the expression $Y_{N}$ becomes
\begin{eqnarray}
Y_{N} \simeq 
2\,\frac{135\,M_{\rm Pl}}{8\pi^3\,1.66}
\sum_{\chi=Z_L,\,Z_R,\,\Phi} 
\dfrac{1}{\sqrt{g_{\rho}}\,g_s}
\left(\dfrac{g_{\chi}\,\Gamma_{\chi\rightarrow\bar{N}N}}{M^2_{\chi}}\right)\,.
\end{eqnarray} 
The above expression of comoving number density for $N$ is similar to
the expression given in \cite{Hall:2009bx}, except the factor $2$, which
is appearing due to the production to two $N$s in the final state
from a single decay of $\chi$.

\end{document}